\documentstyle[12pt]{article}

\newcommand{\gsim}{\mbox{\raisebox{-.3em}{$\stackrel{>}{\sim}$}}}
\newcommand{\lsim}{\mbox{\raisebox{-.3em}{$\stackrel{<}{\sim}$}}}
\renewcommand{\H}{I\!\!H}

\newcommand{\D}{I\!\!D}
\begin{document}
\begin{titlepage}

\title{Quantum cosmology in the energy representation 
         \thanks{
         Work supported by the Austrian Academy of Sciences
         in the framework of the ''Austrian Programme for
         Advanced Research and Technology''.}}

\author{Franz Embacher\\
        Institut f\"ur Theoretische Physik\\
        Universit\"at Wien\\
        Boltzmanngasse 5\\
        A-1090 Wien\\
        \\
        E-mail: fe@pap.univie.ac.at\\
        \\
        UWThPh-1996-32\\
        gr-qc/9605021} 
\date{}

\maketitle

\begin{abstract}
The Hawking minisuperspace model (closed FRW geometry with a 
homogeneous massive scalar field) provides a fairly non-trivial 
testing ground for fundamental problems in quantum cosmology. 
We provide evidence that the Wheeler-DeWitt equation admits a 
basis of solutions that is distinguished by analyticity properities 
in a large scale factor expansion. As a consequence, the space of 
solutions decomposes in a preferred way into two Hilbert spaces 
with positive and negative definite scalar product, respectively. 
These results may be viewed as a hint for a deeper significance of 
analyticity. If a similar structure exists in full 
(non-minisuperspace) models as well, severe implications on the 
foundations of quantum cosmology are to be expected. 
\par
Semiclassically, the elements of the preferred basis describe
contracting and expanding universes with a prescribed value 
of the matter (scalar field) energy. Half of the basis elements 
have previously been constructed by Hawking and Page in a wormhole 
context, and they appear in a new light here. The technical tools 
to arrive at these conclusions are transformations of the wave 
function into several alternative representations that are based 
on the harmonic oscillator form of the matter energy operator, and 
that are called oscillator, energy and Fock representation. The 
framework defined by these may be of some help in analyzing the 
Wheeler-DeWitt equation for other purposes as well. 
\medskip

\end{abstract}

\end{titlepage}

\section{Introduction}
\setcounter{equation}{0}

The aim of this article is twofold: In the weaker
sense it provides a formulation of minisuperspace
quantum cosmology with a massive scalar field
(the so-called Hawking model) 
in terms of representations that are based on
eigenstates of the matter energy operator $E$. 
Since this object does not commute with the 
operator defining the Wheeler-DeWitt equation, 
its eigenstates do not satisfy the latter, 
but their harmonic oscillator form motivates 
a change of representation for the wave function and
an according transformation of the form of the
Wheeler-DeWitt equation. 
All issues concerning the excitation of the scalar field
oscillator modes become particularly transparent 
in the ''energy representation'' and 
related versions thereof. Also, our formulation may facilitate 
attempts to solve the exact Wheeler-DeWitt equation for 
whatever purpose. In the post-inflationary regime the matter energy
eigenstates can be extended to approximate 
solutions of the Wheeler-DeWitt equation at the level
of the WKB-approximation. 
\medskip

In the stronger sense we present perspectives for the
construction of {\it exact} wave functions that 
coincide with the approximate ones in the WKB-domain. 
Moreover, these states seem to have a 
{\em preferred status}, defined by analyticity properties
in an expansion for large values of the scale factor. 
Although some of our conclusions are only conjectures, 
we seem to be able to define two exact Hilbert spaces 
of wave functions, playing a distinguished role. 
This is in some formal analogy to the one-particle
Hilbert spaces of negative/positive frequencies in the
flat Klein-Gordon equation. In this case
the crucial property enabling us to decompose
the space of solutions into two Hilbert spaces is 
Lorentz invariance. In contrast, the symmetries
and covariance properties of the Wheeler-DeWitt equation 
do not suffice to provide such a decomposition
\cite{DeWitt}. 
What we found is that the asymptotic analyticity structure 
of solutions may play a role analogous to Lorentz
invariance. 
Accepting a preferred decomposition has of course 
implications for the conceptual basis of quantum 
cosmology. What we cannot answer at the moment is the question
to what extent this structure will apply for the 
full (non-minisuperspace) theory. 
\medskip

The basic variables in the Hawking
minisuperspace model 
\cite{Hawking2} 
are the
scale factor $a$ of a closed Friedmann-Robertson-Walker 
universe and the value $\phi$ of a homogeneous,
minimally coupled massive scalar field. 
The external parameters are the mass $m$ of the scalar field
and a numerical constant $p$ representing the operator 
ordering ambiguity in the Wheeler-DeWitt equation. 
Apart from some general remarks, 
we will work entirely within this model.
\medskip

In Section 2, we introduce as tools for the analysis
a representation in terms of the harmonic
oscillator eigenfunctions associated with the 
matter energy, and a suitable Fock space notation.
We have designed this Section as self-contained as 
possible and provided various formulae that are helpful
in dealing with different representations of the
wave function. 
For later reference, we distinguish between position,
oscillator, energy and Fock representation. 
In Section 3, we use the fact that matter energy 
is approximately conserved after inflation to write down 
energy eigenstates that approximately satisfy
the Wheeler-DeWitt equation in the corresponding domain of
minisuperspace. 
These states have been used by other authors as well,
at the level of the WKB-approximation. 
The asymptotic structure for values of $a$ larger than 
classically allowed is related to the classical domain 
by an appropriate WKB matching procedure. 
The same asymptotic structure appears when the semiclassical 
WKB-expansion method is applied straightforwardly. 
\medskip

In Section 4, we write down a fairly general expression for
wave functions in the representation based on the
original variables $a$ and $\phi$. We impose an analyticity
requirement in terms of an expansion in inverse powers of $a$ 
that seems to single out an exact version of the
approximate states considered in Section 3. 
In Section 5, a similar procedure, when applied to the
Wheeler-DeWitt equation in the 
oscillator and Fock representations, seems to generate 
identical results. 
Assuming these to hold, we end up with a basis of 
exact and distinguished wave functions $\Xi_n^\pm(a,\phi)$, 
each describing (at the level of WKB-identification) 
an ensemble of collapsing/expanding universes with
matter energy $(n+\frac{1}{2})m$. 
Half of these wave functions have been constructed 
previously in a wormhole context by
Hawking and Page
\cite{HawkingPage2}, 
and the way they emerge in our framework sheds new
light on them. 
Using the indefinite
Klein-Gordon type scalar product $Q$ on the space
of solutions of the Wheeler-DeWitt equation in Section 6, 
we decompose this space into two
Hilbert spaces, with positive/negative definite scalar 
product, respectively. 
If this decomposition is considered as a
{\it preferred} one, relevant for interpretation,
positive probabilities may be written
down by means of conventional Hilbert space techniques. 
As an example, we expand the no-boundary
wave function in terms of this basis in Section 7. 
Such an expansion may be called ''energy representation'' 
in the sense of representing a state entirely
in terms of expansion coefficients with respect to
a basis of {\it solutions} of the Wheeler-DeWitt equation that
describe universes of definite energy. 
Thereby, also the role of the above-mentioned Hawking-Page
solutions as providing only half of a basis
is illustrated. 
Some concluding remarks, concerning the
tunnelling wave function and the significance of
the structure we found for the conceptual issues of
quantum cosmology are given in Section 8. 
\medskip

To conclude this introduction, we comment on the units used
(see Ref. \cite{Hawking2}). 
Let $\sigma^2=2G/3\pi$. In what follows a tilde shall denote 
''true'' physical quantities in units in which $c=1$. 
The Planck mass and length are $\widetilde{m}_P =(\hbar/G)^{1/2}$ 
and $\widetilde{\ell}_P =\hbar/\widetilde{m}_P$.
The FRW space-time metric is given by
\begin{equation}
ds^2 =  - \widetilde{N}(t)^2 dt^2 +
\widetilde{a}(t)^2 d\sigma_3^2 
\label{FRW1}
\end{equation}
where $d\sigma_3^2$ is the metric on the round unit three-sphere.
The scale factor and lapse are rescaled as 
$\widetilde{a}=\sigma a$ and $\widetilde{N}=\sigma N$. 
Let furthermore be $\widetilde{\phi}$ the (spatially homogeneous)
scalar field (with dimension $\hbar^{1/2}\ell^{-1}$, $\ell$
denoting length) and 
$\widetilde{m}$ its mass, and set 
$\phi = \sigma\pi\sqrt{2}\,\widetilde{\phi}$ and
$m=\sigma\widetilde{m}$. (A general scalar field
potential would be redefined as
$V(\phi)= 2 \pi^2\sigma^4 \widetilde{V}(\widetilde{\phi})$.
For the massive case we have 
$\widetilde{V}(\widetilde{\phi})=
\widetilde{m}^2\widetilde{\phi}^2/2\hbar^2$
and 
$V(\phi)=m^2\phi^2/2\hbar^2$).
According to this scheme we rescale the Planck mass and
length as $m_P = \sigma \widetilde{m}_P=(2 \hbar/3\pi)^{1/2}$ 
and $\ell_P=\sigma^{-1} \widetilde{\ell}_P=(3\pi\hbar/2)^{1/2}$.
The Lagrangian resulting from this ansatz, when expressed in 
terms of the rescaled variables, is thus
\begin{equation}
L =\,\frac{1}{2}\left(-\,\frac{a \dot{a}^2}{N} + N a\right)
+\frac{a^3}{2}\left( \frac{\dot{\phi}^2}{N} -\, 
\frac{N m^2}{\hbar^2} \phi^2 \right),
\label{lagrangian}
\end{equation}
the action being $S=\int dt L$, with dimension $\hbar$. 
From now on, it is easy to restore the original variables at
any stage of the quantization procedure. 
(This will be helpful for the reader who likes to go into the
details of the semiclassical expansion carried out at the
end of Section 3). 
By using units in which $\hbar=1$, the variable $a$
as well as the mass parameter $m$ become dimensionless, and
we have $m_P=(2/3\pi)^{1/2}\approx 0.46$ and
$\ell_P =(3\pi/2)^{1/2}\approx 2.17$. The ratio between
scalar field mass and Planck mass is thus
$\widetilde{m}/\widetilde{m}_P=m/m_P\approx 2.17\, m$. 
In order to account for the necessary amount of
density fluctuations 
\cite{HalliwellHawking}\cite{LindeInfl}, 
we expect the mass parameter to be $m\approx 10^{-6}$, hence 
much smaller than $1$. 
\medskip

\section{Representations for states}
\setcounter{equation}{0}

Let us as preliminaries write down the equations governing the
classical dynamics of the Hawking model
\cite{Page}\cite{Kiefer1}, 
i.e. the trajectories 
$(a(t),\phi(t))$ in the minisuperspace manifold
$\{(a,\phi)|a>0\}$. 
If the space-time lapse $N$ has been fixed (e.g. by
assuming a functional dependence $N\equiv N(a,\phi)$), the
relation between the momenta and the time-derivatives of
the minisuperspace variables is given by
\begin{equation}
p_a = - \,\frac{a}{N}\,\frac{da}{dt}\qquad\qquad 
p_\phi = \frac{a^3}{N}\,\frac{d\phi}{dt}\, .
\label{momhm}
\end{equation}  
In the time gauge $N=1$, 
the classical constraint equation reads
\begin{equation}
\dot{a}^2 + 1 = a^2(\dot{\phi}^2+ m^2\phi^2)\, . 
\label{classconstraint}
\end{equation}
When evaluated at some initial time $t_0$, it may be 
interpreted as a restriction on the set of initial
conditions $(a(t_0),\phi(t_0),\dot{a}(t_0),\dot{\phi}(t_0))$.
Once it is imposed for {\it all} times, the time evolution 
equations reduce to
\begin{equation}
\ddot{\phi} + 3 \,\frac{\dot{a}}{a}\,\dot{\phi} + m^2 \phi = 0\, . 
\label{classevol}
\end{equation}
The corresponding $\ddot{a}$-equation which arises from the
Lagrangian formalism is automatically satisfied for all 
times on account of (\ref{classconstraint}). In a general
time gauge, the above equations apply after replacing $d/dt$ by
$N^{-1}\, d/dt$. 
\medskip

We will not go into the details of the properties of 
classical trajectories. They have been studied by a
number of authors (see e.g. Refs. 
\cite{Page}\cite{Kiefer1}). 
Let us just note that a typical (outgoing) classical trajectory
leaves the inflationary domain
$m a |\phi|\gg 1$, $|\phi|\gg 1$ when $|\phi|$ settles
to the order of unity. The subsequent evolution is of
the matter dominated type, with $\phi$ undergoing
rapid oscillations. Eventually (when the amplitude of
$\phi$ falls into the domain of negative potential,
$m a|\phi|<1$), the scale factor $a$ reaches
its maximum value, and the universe recollapses again.
During this process the energy of the scalar field
\begin{equation}
E = \frac{a^3}{2}\, (\dot{\phi}^2 + m^2 \phi^2) 
\label{Eclassical}
\end{equation}
is approximately conserved. The amplitude of the oscillations
of $\phi$ is given by
\begin{equation}
\phi_{\rm ampl} \approx \frac{\sqrt{2E}}{m a^{3/2}} \, . 
\label{phiampl}
\end{equation}
The maximum scale factor is $a_{\rm max}\approx 2 E$.
(In order not to deal too much with approximate quantities,
one may simply {\em define} $E=a_{\rm max}/2$). 
The value of $a$ at which the trajectory enters (and leaves) the
domain $|\phi|\lsim 1$ (i.e. the value of the
scale factor at the end of inflation) 
is roughly given by $a_{\rm min}\approx (E/m^2)^{1/3}$. 
Hence, only if $a_{\rm min}\ll a_{\rm max}\,$, i.e.
$E\gg m^{-1}$, we have a post-inflationary classical
evolution at all. 
\medskip

Canonical quantization is achieved by rewriting the constraint 
in Hamiltonian form and performing the usual substitutions
(see e.g. \cite{Halliwell3}).  
The result is the minisuperspace Wheeler-DeWitt equation
\begin{equation}
{\cal H}\,\psi = 0 \, , 
\label{wdw}
\end{equation}
where the state is represented as a wave function
$\psi(a,\phi)$. 
Since in what follows different representations will be used, 
we distinguish between the notation of an ''operator''
and its ''representation'', in particular if derivatives with 
respect to $a$ are involved. Let ${\cal D}_a$ be the operator that 
acts on a wave function $\psi(a,\phi)$ as the partial derivative
$\partial_a\,$. Then the Wheeler-DeWitt operator is given by 
\begin{equation}
{\cal H} = 
{\cal D}_a {\cal D}_a
+ \frac{p}{a} \,{\cal D}_a + 2 a E - a^2\, , 
\label{wdwoperator1}
\end{equation}  
where
\begin{equation}
E = -\,\frac{1}{2 a^3} \,\partial_{\phi\phi} +
\frac{1}{2}\, a^3 m^2 \phi^2 
\label{E1}
\end{equation}
represents the total matter energy (cf.
\ref{Eclassical} for its classical analogue), and
$p$ is a parameter accounting for the operator ordering
ambiguity.
There are good arguments in favour of $p=1$ (see e.g. Ref. 
\cite{HawkingPage}), 
but we will leave it unspecified. The last term $-a^2$
in (\ref{wdwoperator1}) represents the spatial curvature. In case
of a spatially flat FRW model, one would omit it, and in case
of an open FRW model one would change its sign. In analogy with
ordinary quantum mechanics, the wave function
$\psi(a,\phi)$ can be said to be in the position representation 
(which is just the defining representation here). 
To be explicit, the Wheeler-DeWitt equation in this representation
reads
\begin{equation}
({\cal H}\psi)(a,\phi) \equiv 
\left(
\partial_{aa} + \frac{p}{a} \,\partial_a
- \,\frac{1}{a^2}\, \partial_{\phi\phi} + m^2 a^4 \phi^2 -a^2
\right) \psi(a,\phi) = 0\, , 
\label{wdwposrep}
\end{equation}
which is the standard form in which it is usually written down 
\cite{Hawking2} 
and which determines its form in all other representations. 
(We will {\it not} make attempts to modify it, e.g. by introducing
different operator orderings or taking square roots of operators
as inspired by the hope of making expressions simple in some 
particular representation. In other words, we are not searching
for an alternative wave equation, but just stick to 
(\ref{wdwposrep}) as the starting point, although written down in 
other representations). 
\medskip

Let us begin our analysis by exploiting the fact that for
any value of $a$ the operator $E$ (as acting on functions of $\phi$)
represents a quantum mechanical harmonic oscillator
with frequency $m$ and mass $a^3$. Its eigenvalues are thus
$E_n = (n+\frac{1}{2}) m$ for non-negative integer $n$. 
Using the combination 
\begin{equation}
\xi = m^{1/2} a^{3/2} \phi \, , 
\label{xi}
\end{equation}
we define an alternative representation of states by 
\begin{equation}
\psi(a,\phi) \equiv m^{1/4} a^{3/4} \, \widehat{\psi}(a,\xi)\, . 
\label{psihat}
\end{equation}
Since $\xi$ plays the role of an oscillator variable, 
one could call $\widehat{\psi}(a,\xi)$ to be in the
oscillator representation. The matter energy operator becomes 
essentially a unit harmonic oscillator, 
\begin{equation}
E = \frac{m}{2} (-\partial_{\xi\xi}+\xi^2) \, . 
\label{E2}
\end{equation}
The operator ${\cal D}_a$ (which was $\partial_a$ in the
representation $\psi(a,\phi)$) takes a different form now.
Let $\D_a$ be the operator that acts on
a wave function in the representation $\widehat{\psi}(a,\xi)$ 
as the partial derivative $\partial_a\,$. Then we have
\begin{equation}
{\cal D}_a = \D_a + \frac{3}{2a}\, {\cal K}
\label{DD}
\end{equation}
with
\begin{equation}
{\cal K} =\frac{1}{2}\,
 \{\phi,\partial_\phi\}\equiv \phi\,\partial_\phi+\frac{1}{2} = 
\frac{1}{2}\,
 \{\xi,\partial_\xi\}\equiv \xi\,\partial_\xi+\frac{1}{2}\, . 
\label{K1}
\end{equation}
The Wheeler-DeWitt operator in the oscillator representation 
$\widehat{\psi}(a,\xi)$ 
is now still given by (\ref{wdwoperator1}), 
but with ${\cal D}_a$ being represented as (\ref{DD}), and 
$\D_a$ being represented as $\partial_a$. 
Thus, one may write
\begin{equation}
{\cal H} = \D_a \D_a + \frac{p}{a}\, \D_a +
\frac{3{\cal K}}{a}\,\D_a + \frac{3(p-1)}{2a^2}\,{\cal K} +
\frac{9}{4a^2}\,{\cal K}^2 + 2 a E - a^2 
\label{wdwoperator2}
\end{equation}
which is valid in {\it any} representation (just as
(\ref{wdwoperator1}) is), as long as by 
$\D_a$ the appropriate operator representation is understood. 
\medskip

The eigenfunctions of $E$ are just those of the unit harmonic
oscillator with coordinate $\xi$. In terms of
Hermite polynomials (which are generated by
$e^{2\xi t-t^2} = \sum_{n=0}^\infty\frac{1}{n!}t^n H_n(t)\,$)  
they read 
\begin{equation}
\Psi_n(\xi) = \frac{H_n(\xi)}{\sqrt[4]{\pi} \sqrt{2^n n!}}\,
e^{-\,\frac{1}{2}\,\xi^2} 
\label{hermite}
\end{equation}
with $n$ a non-negative integer. 
By expansion with respect to these, 
we define a further way of writing wave functions 
\begin{equation}
\widehat{\psi}(a,\xi)=\sum_{n=0}^\infty f_n(a)\Psi_n(\xi)\, , 
\label{fnrep}
\end{equation}
where the component functions $f_n(a)$ may be regarded as
providing the state in the
energy representation (although we will encounter a further 
meaning of this word later on). This notation is justified
by the fact that the operator $E$ is now diagonal:
it sends $f_n(a)$ to $E_n f_n(a)$. Its action may symbolically 
be written as $(E f)_n(a)= E_n f_n(a)$. 
Note that even $n$ belongs to the even 
($\psi(a,-\phi)=\psi(a,\phi)$) 
and odd $n$ to the odd ($\psi(a,-\phi)=-\psi(a,\phi)$) sector of
wave functions. Since the $\Psi_n$ are an orthonormal basis,
(\ref{fnrep}) may be inverted to give the oscillator
excitations 
\begin{equation}
f_n(a)=\int_{-\infty}^\infty d\xi\, \Psi_n(\xi) 
\widehat{\psi}(a,\xi) 
\label{fninv}
\end{equation}
in terms of the wave function in the oscillator representation. 
\medskip

In performing (\ref{fnrep}) we have implicitly assumed that the 
wave function $\psi$ 
is sufficiently well-behaved for large $\xi$ (or $\phi$) so as to 
allow for such an expansion. A quite restrictive condition
on general wave functions would be square integrability in the matter
variable $\xi$ (or $\phi$, which is equivalent). Although
this would offer a Hilbert space structure for any value of $a$, 
the more interesting candidate wave functions of the universe
are of distributional character with respect to this structure,
and we just assume that the expansion (\ref{fnrep}) is possible.
The formal squared norm of wave functions 
\begin{equation}
\int_{-\infty}^\infty d\phi\,\psi^*(a,\phi) \psi(a,\phi) =
\int_{-\infty}^\infty d\xi\,
\widehat{\psi}^*(a,\xi) \widehat{\psi}(a,\xi)=
\sum_{n=0}^\infty f_n^*(a) f_n(a)
\label{norm}
\end{equation}
may but need not be finite. The prefactor $m^{1/4} a^{3/4}$
in (\ref{psihat}) has been chosen so as to give these
expressions a simple form. The formal hermiticity of operators
like $i\partial_\phi$, $i\partial_\xi$, $i{\cal K}$ and $E$ with 
respect to the according scalar product is evident. 
\medskip

The Wheeler-DeWitt operator in this representation
is given by (\ref{wdwoperator2})
with $\D_a=\partial_a$, $E$ acting diagonal with eigenvalues
$E_n$ and ${\cal K}$ acting as (in a symbolic notation) 
\begin{equation}
({\cal K} f)_n(a)=-\,\frac{1}{2}\sqrt{n(n-1)}\, f_{n-2}(a) +
\frac{1}{2} \sqrt{(n+1)(n+2)}\, f_{n+2}(a) \, , 
\label{Kact}
\end{equation}
its square being given by 
\begin{eqnarray}
({\cal K}^2 f)_n(a)&=& 
\frac{1}{4}\sqrt{n(n-1)(n-2)(n-3)}\, f_{n-4}(a)-  \nonumber\\
& & \frac{1}{4}\Big(n(n-1)+(n+1)(n+2)\Big)\, f_n(a) +
\label{Kactsquare}\\
& &\frac{1}{4} \sqrt{(n+1)(n+2)(n+3)(n+4)}\, f_{n+4}(a) \, .
\nonumber 
\end{eqnarray}
For completeness, we write down the Wheeler-DeWitt equation
in this representation explicitly: 
\begin{eqnarray}
({\cal H}f)_n(a) &\equiv& 
\Big( \partial_{aa}+\frac{p}{a}\,\partial_a \Big) f_n(a)
+\frac{3}{a}\,\partial_a ({\cal K} f)_n(a) +
\frac{3(p-1)}{2a^2}\,({\cal K} f)_n(a) +\nonumber\\
& &\frac{9}{4a^2}\, ({\cal K}^2 f)_n(a) +
\Big( 2 a E_n - a^2 \Big) f_n(a) = 0\, . 
\label{wdwfn}
\end{eqnarray}
It involves differences with respect to $n$ rather than derivatives.
Note that the highest component function $f_{n+4}(a)$ appears 
algebraically. Hence, the Wheeler-DeWitt equation 
in the energy representation is just a recursive expression
for $f_n(a)$ ($n\ge 4$) in terms of $(f_0(a),\dots,f_3(a))$. 
\medskip

The form (\ref{E2}) of $E$ amounts to define formal annihilation
and creation operators
\begin{eqnarray}
{\cal A} &=& a^{3/2}\,\sqrt{\frac{m}{2}}\,\phi + 
\frac{1}{a^{3/2}\sqrt{2m}}\, \partial_\phi \ =
\frac{1}{\sqrt{2}}(\xi + \partial_\xi)
\label{anni}\\
{\cal A}^\dagger &=& a^{3/2}\,\sqrt{\frac{m}{2}}\,\phi - 
\frac{1}{a^{3/2}\sqrt{2m}}\, \partial_\phi \ =
\frac{1}{\sqrt{2}}(\xi - \partial_\xi)
\label{crea}
\end{eqnarray}
from which 
\begin{equation}
E = m \left( {\cal A}^\dagger {\cal A} + \frac{1}{2}\right)
\equiv m \left({\cal N} + \frac{1}{2}\right)
\label{E3}
\end{equation}
and
\begin{equation}
{\cal K} = \frac{1}{2} \left(
{\cal A}^2 - ({\cal A}^\dagger)^2 \right) \, . 
\label{K2}
\end{equation}
The square of the latter turns out
to be 
\begin{equation}
{\cal K}^2=\frac{1}{4}\left(
{\cal A}^4 -{\cal N}({\cal N}-1)- ({\cal N}+1)({\cal N}+2) 
+ ({\cal A}^\dagger)^4   \right)\, . 
\label{Ksquare}
\end{equation} 
The index $n$ of the components $f_n(a)$ represents the
eigenvalues of the oscillator number operator 
${\cal N}={\cal A}^\dagger {\cal A}$. 
The dagger denotes hermitean conjugation in a formal sense,
with respect to the scalar product associated with the
squared norm expressions (\ref{norm}). 
By formally identifying
$\Psi_n(\xi)$ with the abstract state $|n\rangle$
and using the commutator relation  
$[ {\cal A},{\cal A}^\dagger\,]=1$, 
we find the usual structure defining a Fock space 
\begin{equation}
{\cal A}|n\rangle=\sqrt{n}\,|n-1\rangle\qquad\qquad
{\cal A}^\dagger|n\rangle=\sqrt{n+1}\,|n+1\rangle \, . 
\label{AA}
\end{equation}
The $n$-the eigenstate is generated out of the ground state
(note that ${\cal A}|0\rangle=0$) by
\begin{equation}
|n\rangle=\frac{({\cal A}^\dagger)^n}{\sqrt{n!}}\,
|0\rangle \, . 
\label{nthstate}
\end{equation}
A given wave function may be written in the form
\begin{equation}
|\psi,a\rangle =\sum_{n=0}^\infty f_n(a) |n\rangle = 
\sum_{n=0}^\infty f_n(a) 
\frac{({\cal A}^\dagger)^n}{\sqrt{n!}}\,
|0\rangle\equiv {\cal F}(a,{\cal A}^\dagger)|0\rangle\, ,  
\label{fockrep}
\end{equation}
thus defining a ''Fock representation'' for states. If
$\langle\psi,a|$ is defined as $\sum_n \langle n|f_n^*(a)$,
or equivalently as $\langle 0|{\cal F}^*(a,{\cal A})$, the
formal squared norm (\ref{norm}) is given by 
$\langle\psi,a|\psi,a\rangle$. 
The analogue of (\ref{fninv}) is
\begin{equation}
f_n(a) = \langle n|\psi,a\rangle\, , 
\label{fnfock}
\end{equation}
and the orthonormality of the oscillator basis carries 
over to $\langle r|s\rangle=\delta_{rs}\,$. 
If $|\psi,a\rangle$ is written as 
${\cal F}(a,{\cal A}^\dagger)|0\rangle$, the operator 
${\cal A}$ is formally represented as 
$\partial/\partial {\cal A}^\dagger$,
whereas ${\cal A}^\dagger$ may be considered as a multiplication 
operator, and $\D_a$ is the partial derivative $\partial_a$. 
The relation between $\widehat{\psi}(a,\xi)$ and
${\cal F}(a,{\cal A}^\dagger)$ has so far been given only through 
a number of intermediate steps, involving Hermite polynomials
and an infinite sum. It can be made more explicit by noting the
identifications of functions of $\xi$ with abstract 
Fock space states
\begin{equation}
e^{i k \xi}\equiv\sqrt{2\pi}\,\sum_{n=0}^\infty 
i^n \,\Psi_n(k)\Psi_n(\xi)
\longleftrightarrow 
\sqrt[4]{\pi}\sqrt{2}\,
e^{-\,\frac{1}{2} k^2}
e^{i\sqrt{2}\, k {\cal A}^\dagger+\frac{1}{2}({\cal A}^\dagger)^2}
|0\rangle\, , 
\label{eikxi}
\end{equation}
for fixed $k$ and 
\begin{equation}
\delta(\xi-\eta)\equiv\sum_{n=0}^\infty 
\Psi_n(\eta)\Psi_n(\xi)
\longleftrightarrow 
\frac{1}{\sqrt[4]{\pi}}\, 
e^{-\,\frac{1}{2}\eta^2} 
e^{\sqrt{2}\,\eta{\cal A}^\dagger-\,\frac{1}{2}({\cal A}^\dagger)^2} 
|0\rangle
\label{deltaxime}
\end{equation}
for fixed $\eta$. 
A further interesting relation illustrating the appearence 
of Hermite polynomials is 
\begin{equation}
\xi^n\longleftrightarrow 
\sqrt[4]{\pi}\sqrt{2}\, (-i)^n\, 2^{-n/2}\, H_n(i{\cal A}^\dagger)
\,e^{\frac{1}{2} ({\cal A}^\dagger)^2}
|0\rangle\, . 
\label{xiq}
\end{equation}
for non-negative integer $n$. 
Using (\ref{deltaxime}), one finds for a given wave function 
$\widehat{\psi}(a,\xi)$ that 
\begin{equation}
{\cal F}(a,{\cal A}^\dagger) = 
\frac{1}{\sqrt[4]{\pi}}
\int_{-\infty}^\infty d\eta\, \widehat{\psi}(a,\eta) 
e^{-\,\frac{1}{2}(\eta-\sqrt{2}\,{\cal A}^\dagger)^2}
e^{\frac{1}{2}({\cal A}^\dagger)^2} \, . 
\label{calF}
\end{equation}
Inverting this relation is not that straightforward. Defining
$\widetilde{\psi}(a,k)$ to be the inverse Fourier transform of
$\widehat{\psi}(a,\xi)$, i.e. 
\begin{equation}
\widehat{\psi}(a,\xi) = \int_{-\infty}^\infty 
\frac{dk}{\sqrt{2\pi}}\, \widetilde{\psi}(a,k) e^{i k \xi}\, , 
\label{Fou}
\end{equation}
and making use of (\ref{eikxi}), equation (\ref{calF}) may be 
expressed alternatively as 
\begin{equation}
{\cal F}(a,{\cal A}^\dagger) = 
\sqrt[4]{\pi}\sqrt{2}
\int_{-\infty}^\infty \frac{dk}{\sqrt{2\pi}} \, 
\widetilde{\psi}(a,k)
e^{-\,\frac{1}{2}(k-i\sqrt{2}\,{\cal A}^\dagger)^2}
e^{-\,\frac{1}{2}({\cal A}^\dagger)^2} \, . 
\label{calFFour}
\end{equation}
This may be inverted to give
\begin{equation}
\widetilde{\psi}(a,k) = \frac{1}{\sqrt[4]{\pi}}
\int_{-\infty}^\infty \frac{d{\cal A}^\dagger}{\sqrt{2\pi}} \, 
{\cal F}(a,{\cal A}^\dagger)
e^{-\,\frac{1}{2}({\cal A}^\dagger+ i\sqrt{2}\,k)^2}
e^{-\,\frac{1}{2} k^2} \, . 
\label{Fourier}
\end{equation}
Combining this last equation with (\ref{Fou}) and interchanging the 
${\cal A}^\dagger$ and $k$ integrations gives a direct formula for
$\widehat{\psi}(a,\xi)$ in terms of ${\cal F}(a,{\cal A}^\dagger)$,
but with an integrand that exists only in a distributional sense.
\medskip

The Wheeler-DeWitt operator in this representation is given
by (\ref{wdwoperator2}) with $\D_a=\partial_a$, $E$ from (\ref{E3}) 
and $\cal K$ from (\ref{K2}). 
Note that, due to (\ref{Ksquare}), the operator ${\cal K}^2$ contains
${\cal A}^4$. This makes the Wheeler-DeWitt equation in the
Fock representation a fourth order 
differential equation for ${\cal F}(a,{\cal A}^\dagger)$. 
\medskip

We are thus able to represent states in various forms, namely as
$\psi(a,\phi)$, $\widehat{\psi}(a,\xi)$, $f_n(a)$, $|\psi,a\rangle$
and ${\cal F}(a,{\cal A}^\dagger)$. The latter three forms are
of course closely related to each other, referring to the
scalar field energy as a variable. The different versions of the 
Wheeler-DeWitt equation are equivalent, which makes the choice of 
representation a matter of convenience and taste.  
\medskip

\section{Approximate solutions}
\setcounter{equation}{0}

The framework of representations as given in the last Section 
is ''kinematic'' in nature, i.e. it makes no reference to the 
Wheeler-DeWitt equation. 
It is clear that an object like $|n\rangle$, for some fixed $n$,
does not even approximately satisfy it.  
On the other hand, we know that classically the energy $E$ is
approximately conserved in the post-inflationary regime. 
We thus expect that there are solutions of the 
Wheeler-DeWitt equation which behave to some accuracy like
\begin{equation}
|\psi,a\rangle=h(a)|n\rangle
\label{gan}
\end{equation}
for values of $a$ which
belong to the post-inflationary classical domain (i.e.
between end of inflation at
$a_{\rm min}\approx(E_n/m^2)^{1/3}$ and maximum size
at $a_{\rm max}\approx 2 E_n$). 
\medskip

Inserting (\ref{gan}) into the Wheeler-DeWitt equation in the
Fock representation of (\ref{wdwoperator2}), the reason why it
cannot be an exact solution turns out to be that $\cal K$ and 
${\cal K}^2$ as represented by (\ref{K2}) and
(\ref{Ksquare}) contain powers of $\cal A$ and
${\cal A}^\dagger$ that mix the elements $|r\rangle$ of the
oscillator basis. These terms arise from the $a$-dependence of 
$E$ in (\ref{E1}). When $a$ is large, they can be neglected, 
and the harmonic oscillator dynamics follows the dynamics
of the gravitational sector. 
This leads to the adiabatic approximation, a technique which
is frequently applied (see e.g. Ref.
\cite{Kiefer1}). 
In order to have a non-trivial classical domain at all
($a_{\rm min}\ll a_{\rm max}$) we must choose 
$n\gg m^{-2}\approx 10^{12}$, hence much larger than $1$. 
Neclecting all non-trivial powers of $\cal A$ and 
${\cal A}^\dagger$ in (\ref{wdwoperator2}), we pick up the 
effective potential term $-\frac{9}{8 a^2}(n^2+n+1)$ from 
${\cal K}^2$. Taking into account the operator ordering 
term as well, we end up with an approximate
equation for the range $a\gg a_{\rm min}$
\begin{equation}
\Big(\partial_{aa}+\frac{p}{a}\,\partial_a
+ U(a) \Big) h(a) = 0 
\label{wdwg}
\end{equation}
with
\begin{equation}
U(a) = 2 a E_n - a^2 -\frac{9}{8a^2}(n^2+n+1)\, . 
\label{Ua}
\end{equation}
From the point of view of the oscillator
basis, this is just the projection of the Wheeler-DeWitt
equation with (\ref{gan}) inserted onto the $n$-the basis
element. Hence, (\ref{wdwg}) is identical to
$\langle n|{\cal H}\,h(a)|n\rangle = 0$. It is the best
we can do when all oscillators other than $|n\rangle$
are ignored. 
\medskip

By redefining $h(a)= a^{-p/2} g(a)$, one kills the
first order derivative. 
In the range $a\gg a_{\rm min}\,$, the $a^{-2}$ contribution 
to $U(a)$ (together with an additional term from
the above redefinition) is small. Neglecting it, the new
approximate equation reads
\begin{equation}
\Big(\partial_{aa}+ 2 a E_n - a^2 \Big) g(a) = 0\, . 
\label{wdwgg}
\end{equation}
The solutions thereof may be
expressed in terms of parabolic cylinder functions
\cite{GradshteynRyzhik}. 
Since this is not very instructive, let us look at some
possible ranges of $a$.
For $a\ll a_{\rm max}\,$, the potential in (\ref{wdwgg}) 
is dominated by $2 a E_n$, the according solutions 
being the two Airy functions
${\rm Ai}(-(2E_n)^{1/3}\,a)$ and ${\rm Bi}(-(2E_n)^{1/3}\,a)$. 
Since the arguments $-(2E_n)^{1/3}\,a$ of these functions
are much less than $-1$, we are in the range in which they 
oscillate rapidly. Up to a multiplicative constant,
the according asymptotic expressions for $g(a)$ are
\begin{equation}
a^{-1/4} \,{\rm osc}
\left(\frac{2}{3}\,(2E_n)^{1/2} a^{3/2}+\frac{\pi}{4}\right)
\label{gsol1}
\end{equation}
with ${\rm osc}=\cos$ for ${\rm Bi}$ and ${\rm osc}=\sin$ for 
${\rm Ai}$. 
\medskip

When $a$ approaches $a_{\rm max}\,$, the $-a^2$ contribution
in the potential becomes important and will slow down the
oscillations. Here we note that so far it was not necessary to
restrict $a$ to be less than $a_{\rm max}\,$. At
$a\approx a_{\rm max}\approx 2 E_n$, the potential vanishes. 
For $a\gg a_{\rm max}\,$, the overall behaviour is exponential,
and we find two asymptotic solutions for $g(a)$ of the type
\begin{equation}
a^{-1/2\,\mp\, E_n^2/2} 
\exp\left(\pm \frac{1}{2}a^2\mp E_n a\right)\, . 
\label{gsol2}
\end{equation}
For given $E_n$, this range is ''classically forbiden'',
because all classical universes with matter energy $E_n$
cannot extend to such sizes. Note that 
for $n\lsim m^{-2}$, the asymptotic solutions (\ref{gsol2}) 
still exist, athough they do not have an oscillatory domain
of the type (\ref{gsol1}). 
\medskip

So far we have not stated anything about the allowed
range of $\phi$. The classical amplitude of oscillations
is given by (\ref{phiampl}), and we expect the approximation
for the oscillating regime (\ref{gsol1}) to hold
if $|\phi|\lsim \phi_{\rm ampl}\,$. In terms of $\xi$,
this means $|\xi| \lsim\sqrt{2n}$. This is in fact just the 
domain in which the oscillator basis $\Psi_n(\xi)$ is considerably
non-zero. (Note that the amplitude of the classical
oscillations with energy $m n$ corresponding to the
Hamiltonian (\ref{E2}) is just $\xi_{\rm ampl}=\sqrt{2n}\,$). 
In the exponential regime (\ref{gsol2}), an 
identical condition on $\xi$ holds. Hence, for $a\gg a_{\rm min}$
the interesting values are concentrated in a stripe around the
$\phi=0$ axis that gets arbitrarily narrow as $a$ increases. 
\medskip

In order to link the asymptotic form (\ref{gsol2}) with
the oscillating behaviour (\ref{gsol1}), we may apply the
standard WKB matching procedure between domains in which the
potential has different sign
\cite{Bialynickietal}. 
Introducing the number 
\begin{equation}
\Theta_n = \frac{\pi}{2}\,E_n^2+\frac{\pi}{4}\, , 
\label{Theta}
\end{equation}
we fix the normalization in the oscillatory region by
defining two real approximate solutions of the type
(\ref{gsol1}) 
\begin{eqnarray}
|\Omega_n^+,a\rangle &=& 
a^{-p/2-1/4}\, (2E_n)^{-1/4}\, 
\cos\left(\Theta_n - \frac{2}{3}\,(2E_n)^{1/2} a^{3/2} \right)
|n\rangle 
\label{omegaplusosc}\\
|\Omega_n^-,a\rangle &=& 
a^{-p/2-1/4}\, (2E_n)^{-1/4}\, 
\sin\left(\Theta_n - \frac{2}{3}\,(2E_n)^{1/2} a^{3/2} \right) 
|n\rangle \, . 
\label{omegaminusosc}
\end{eqnarray}
The result of applying the WKB matching procedure 
(which is too boring to be shown here) 
is that, in the exponential domain, 
\begin{eqnarray}
|\Omega_n^+,a\rangle &=& 
a^{-p/2-1/2} \left(\frac{2a}{E_n}\right)^{-E_n^2/2} 
\exp\left(\frac{1}{2}a^2 - E_n a + \frac{1}{4}E_n^2\right)
|n\rangle 
\label{omegaplusexp}\\
|\Omega_n^-,a\rangle &=& \frac{1}{2}\,
a^{-p/2-1/2} \left(\frac{2a}{E_n}\right)^{E_n^2/2} 
\exp\left(-\,\frac{1}{2}a^2 + E_n a - \frac{1}{4}E_n^2\right)
|n\rangle \, . 
\label{omegaminusexp}
\end{eqnarray}
For $n\lsim m^{-2}$ there is no oscillatory region, and we
define the asymptotic normalization by the these two 
expressions as well.
The factor $\frac{1}{2}$ in (\ref{omegaminusexp}) 
may look a bit stange, but it is an immediate consequence
of the matching procedure (it is actually part of the standard 
formulae) and the fact that we have arranged the normalization
of the oscillating behaviour 
(\ref{omegaplusosc})--(\ref{omegaminusosc}) in a symmetric way. 
Note that the Airy functions ${\rm Ai}(x)$ and ${\rm Bi}(x)$, 
when expanded for 
$x\rightarrow -\infty$ and $x\rightarrow \infty$ (i.e. in the
oscillatory and in the exponential domain) display a 
precisely analogous factor $\frac{1}{2}$ for ${\rm Ai}(x)$ 
(cf. also (\ref{gsol1})). 
\medskip

Defining two other sets of approximate solutions $\Xi_n^\pm$ by
\begin{equation}
\Xi_n^\pm = e^{\pm i \Theta_n}
(\Omega_n^+ \mp i \Omega_n^-) \, , 
\label{Xidef}
\end{equation}
or, inversely, 
\begin{eqnarray}
\Omega_n^+ &=& \frac{1}{2}\left(e^{-i\Theta_n}\,\Xi_n^+ +
e^{i\Theta_n}\,\Xi_n^-\right)
\label{OmXi1}\\
\Omega_n^- &=& \frac{1}{2}\left(e^{-i\Theta_n}\,\Xi_n^+ - 
e^{i\Theta_n}\,\Xi_n^-\right)\, , 
\label{OmXi2}
\end{eqnarray}
we find, in the oscillatory regime,
\begin{equation}
|\Xi_n^\pm\rangle = a^{-p/2-1/4} (2E_n)^{-1/4}\, 
\exp\left(\pm\frac{2i}{3} \,(2E_n)^{1/2} a^{3/2} \right) 
|n\rangle \, . 
\label{Xiosc}
\end{equation}
The involved structure of prefactors was necessary in order
to achieve this simple form. 
The significance of the normalization of this expression 
(in particular the factor $(2E_n)^{-1/4}$) will
become clear later. 
\medskip

Thus, we have a set of approximate solutions
at hand that correspond to 
prescribed values of the matter energy. 
These functions have been used by Kiefer in his
discussion of wave packets in the Hawking model 
\cite{Kiefer1}.
Also, the exponential behaviour of $\Omega_0^+(a,\phi)$ 
has been displayed by Page
\cite{Page} 
as the dominant part of the no-boundary wave function for
small $\phi$ and large $a$. 
The $\Omega_n^-$ also occur in the work of Hawking and
Page 
\cite{HawkingPage2}, 
to which we will refer later on. 
\medskip

The physical significance of any of the wave functions
$\Xi_n^\pm$ with $n\gg m^{-2}$ 
at the WKB-level is to represent an ensemble of
contracting/expanding classical universes with 
post-inflationary matter energy $E_n$. 
The expressions (\ref{Xiosc}) may be viewed as
semiclassical WKB-states around a family of expanding classical
backgrounds with action 
$S_0(a)= -\,\frac{2}{3}\sqrt{2E} a^{3/2}$ (this is discussed
in a bit more detail in Ref. 
\cite{FE8}). 
Note however that the classical matter energy $E$ is 
approximately conserved {\it only} in the domain $|\phi|\ll 1$, 
i.e. ''after'' inflation. Hence, the individual 
$|\Xi_n^\pm, a\rangle$ can be expected to contain essentially
a {\it single} oscillator excitation 
(i.e. of $|n\rangle$) {\it only} if 
$a\gg a_{\rm min}\approx (E_n/m^2)^{1/3}$. For
smaller values of $a$ we expect $|\Xi_n^\pm, a\rangle$ 
to be a non-trivial superposition of (virtually)
{\it all} oscillators $|r\rangle$ and thus all approximate 
expressions we gave for these wave functions to break down. 
Furthermore it is likely that these states are not
well-behaved for small $a$. This is because there are many
classical trajectories to some given value of
$a_{\rm max}$ which behave quite singular as 
$a\rightarrow 0$ (namely of the ''collapse'' type
$|\phi|\rightarrow\infty$).
Also, we cannot expect $|\Xi_n^\pm, a\rangle$ to satisfy
nice properties near the zero potential curve
$a^2m^2\phi^2=1$, along which usually ''nucleation''
is assumed to occur. 
In the case $n\lsim m^{-2}$ we can talk about pure tunnelling 
states that do not correspond to classical universes at all. 
\medskip 

Although, so far, the wave functions have only been identified 
as approximate solutions of the Wheeler-DeWitt equation, 
it is reasonable to suppose the existence of {\it exact} 
solutions that behave qualitatively in the same way. 
In the range $a\gg a_{\rm min}$ this means that
the excitations of oscillators other than $|n\rangle$
may be non-zero but are small. 
Moreover, we expect to obtain a {\it basis} for the
set of solutions: At some fixed value of $a$, the degrees 
of freedom contained in $\{\Omega_n^\pm\}$ correspond to two free 
functions. Hence, at the approximate level, any initial data
$(\psi,\partial_a \psi)|_{a=a_{\rm ini}}$
may be expanded in terms of the oscillators 
$|n\rangle$ (i.e. in terms of the functions 
$\Psi_n(m^{1/2} a^{3/2} \phi)\equiv\Psi_n(\xi)$). 
This behaviour is expected to carry over to the exact case. 
\medskip 

For large $a$, the oscillator mixing operators
(powers of ${\cal A}$ and ${\cal A}^\dagger$) 
contained in ${\cal K}$ and ${\cal K}^2$ get
suppressed. Hence we expect the approximation of
the predominance of {\it one} oscillator $|n\rangle$ in
a wave function to 
become arbitrarily accurate as $a\rightarrow\infty$. In this
limit, we may treat all states belonging to 
non-negative integers $n$ at the same footing (including the pure 
tunnelling states $\Omega_n^\pm$ for small $n$). 
One must of course be aware that an exactification of 
our wave functions will leave a considerable amount of
freedom. However, by combining the large-$a$ expansion
with the assumption of certain analyticity properties, we will
encounter an appearently 
{\it distinguished} way to single out a unique exact solution
for any choice of the $\pm$ label and any $n$. 
\medskip

Let us at the end of this Section provide another argument 
leading to the same exponential behaviour of wave functions.
One may apply the standard semiclassical WKB-expansion scheme 
\cite{Kiefersemi},
based on a Born-Oppenheimer approximation, 
by rewriting the Wheeler-DeWitt equation 
(\ref{wdwposrep}) in terms of units that make the Planck length 
explicit. At the end of the introduction we have displayed the 
relation between the ''true'' quantities $\widetilde{a}$,
$\widetilde{\phi}$, $\widetilde{m}$ and the rescaled ones
$a$, $\phi$ and $m$. Restoring the true units, one finds
the kinetic part of the gravitational field multiplied by the
Planck length squared and the curvature contribution multiplied
by the inverse of the Planck length squared. The matter sector
remains unaffected (up to a numerical factor of order unity). This 
structure may equivalently be written down
by introducing a formal book-keeping parameter
$\lambda$ (playing the role of the Planck length) 
that is treated as a small quantity (and reset equal to
$1$ in the end). The Wheeler-DeWitt equation (\ref{wdwposrep})
thus becomes
\begin{equation}
\left( \lambda^2 \Big(
\partial_{aa} + \frac{p}{a} \,\partial_a \Big)
- \,\frac{1}{a^2}\, \partial_{\phi\phi} + m^2 a^4 \phi^2 
-\frac{a^2}{\lambda^2} 
\right) \psi(a,\phi) = 0\, . 
\label{wdwlambda}
\end{equation}
This may formally be achieved by replacing 
$a\rightarrow \lambda^{-1} a$,
$\phi\rightarrow\lambda\phi$ and
$m\rightarrow\lambda m$ in (\ref{wdwposrep}). 
According to the WKB-description we expand a 
wave function as 
\begin{equation}
\psi(a,\phi) = \exp\left( i\Big(
\frac{S_0(a,\phi)}{\lambda^2}+S_1(a,\phi)+
\lambda^2 S_2(a,\phi) + \lambda^4 S_3(a,\phi)+\dots\Big)\right) \, . 
\label{psilambda}
\end{equation}
This is the direct (''naive'') semiclassical treatment of the
Hawking model. It essentially assumes that the gravitational field
variable $a$ is ''(quasi)classical'', whereas the matter field $\phi$
plays the full quantum role. 
Inserting the ansatz (\ref{psilambda}) into (\ref{wdwlambda}) and 
isolating powers of $\lambda^2$ yields a sequence of equations for the
$S_j(a,\phi)$. In the end, we restore the original
Wheeler-DeWitt equation by setting $\lambda=1$. 
\medskip

The equation at $O(\lambda^{-4})$ turns out to be just
$\partial_\phi S_0(a,\phi)=0$, telling us that
$S_0\equiv S_0(a)$. 
At $O(\lambda^{-2})$ we obtain 
the equation $(\partial_a S_0(a))^2=-a^2$. 
Formally, this is the Hamilton-Jacobi equation for the pure 
gravitational field and thus fits into the general semiclassical
scheme. 
On the other hand, it has no real solution, which just reflects the
fact that an empty closed FRW universe does not exist.
Proceeding straightforwardly, we write down the two imaginary
solutions $S_0(a)=\mp i a^2/2$. The WKB-phase factor
$e^{i S_0(a)}\equiv e^{\pm a^2/2}$ thus provides already 
the dominant large-$a$ behaviour of $\Omega_n^\pm(a,\phi)$.
Rescaling the next order contribution as
$e^{i S_1(a,\phi)}=a^{-1/2-p/2} \chi(a,\phi)$, we find
at $O(\lambda^0)$ the Wheeler-DeWitt equation to state 
\begin{equation}
(\pm \partial_a + E) \chi(a,\phi) = 0
\label{euclSchr}
\end{equation}
with $E$ from (\ref{E1}). This is the step where usually
the effective Schr{\"o}dinger equation (as a minisuperspace
version of the Tomonaga-Schwinger equation) arises 
\cite{Kiefersemi}. 
Obviously, we encounter a Euclidean type Schr{\"o}dinger equation.
In the adiabatic approximation, one neglects the $a$-dependence of
$E$. By introducing the oscillator variable $\xi$ from 
(\ref{xi}), the 
energy operator becomes represented as (\ref{E2}). The adiabatic
approximation amounts to keep the derivative $\partial_a$
unchanged in (\ref{euclSchr}). Thus, factorizing 
$\widehat{\chi}\equiv m^{-1/4}a^{-3/4}\chi$ 
(cf. (\ref{psihat})) into
a product $A(a) X(\xi)$ (in the usual context this would be a
stationary state) yields immediately an eigenvalue
equation, hence $X(\xi)=\Psi_n(\xi)$ for some non-negative
integer $n$, as well as the ''time evolution'' prefactor
$A(a)=e^{\mp E_n a}$. Putting everything together, we
have reproduced qualitatively the behaviour
(\ref{omegaplusexp})--(\ref{omegaminusexp})   
of the wave functions $\Omega_n^\pm\,$. 
The additional factors of the type $a^{\mp E_n^2/2}$ 
are effects at $O(\lambda^2)$. 
\medskip

Thus, for large $a$, the approximate
solutions $\Omega_n^\pm(a,\phi)$ may be considered 
as semiclassical states, 
built around a ''pure tunneling'' background gravitational field
(which nevertheless --- in the way how it appears in a 
WKB-expansion --- displays formal similarities to a
true classical background). This way of looking at things
may seem a bit strange, but since we are faced with a wave
equation (as opposed to classical equations of motion)
we cannot exclude that the tunneling region $a\gg a_{\rm max}$
plays an important role in the structure of the space of solutions 
or in the conceptual foundations of a quantum theory of the
universe. 
This may be in some correspondence with 
the idea of a Euclidean path-integral 
\cite{Hawking1}\cite{HartleHawking} 
or some other principle which provides some additional structure 
that is invisible for the semiclassical techniques. 
Anyway, it may be taken as a further motivation in favour of 
examining the limit $a\rightarrow\infty$ of the Wheeler-DeWitt 
equation. 
\medskip

\section{Exact solutions in the position representation}
\setcounter{equation}{0}

Leaving the level of approximate wave functions, we will try now
to define a large-$a$ expansion scheme that enables us to
specify a set of exact solutions of the Wheeler-DeWitt equation.
An appropriate ansatz is modelled
according to the 
asymptotic behaviour of $\Omega_n^\pm$ as given by
(\ref{gsol2}), multiplied by $|n\rangle$, i.e. by the
function $\Psi_n(\xi)$. The dominant behavour for large $a$ and
fixed $\phi$ of the latter is given by the exponential
$e^{-\xi^2/2}\equiv e^{- m a^3\phi^2/2}$ (see equations
(\ref{xi}) and (\ref{hermite})). This is followed by the
factor $e^{\pm a^2/2}$. Hence, in a large-$a$ expansion
at fixed $\phi$, the leading order in the exponential
is $a^3$, with a $\phi$-dependent coefficient. 
Separating the positive powers of $a$, as appearing in the
exponential, from a pure $a^q$ term (which may represent
$e^{q \ln a}$) and 
expanding the remainder in terms of $a^{-1}$, 
we write down as a general ansatz in the position
representation 
\begin{equation}
\psi(a,\phi)= a^q\,\exp\left( 
F_3(\phi)a^3+F_2(\phi)a^2+F_1(\phi)a \right) 
\sum_{r=0}^\infty \frac{G_r(\phi)}{a^r} \, . 
\label{generalansatz}
\end{equation}
This looks fairly general, the only severe restriction being that
only {\em one} exponential of the above type is involved. 
In a paper on wormhole solutions, Hawking and Page  
\cite{HawkingPage2} 
have effectively used a similar type of expansion, and some 
of the structure we will encounter appears there as well. 
By inserting the
ansatz into the Wheeler-DeWitt equation (\ref{wdwposrep}) and
separating powers of $a$, one obtains a sequence of ordinary
differential equations for the functions $F_r(\phi)$ and
$G_r(\phi)$, thereby expecting the freedom of choosing arbitrary 
integration constants.
\medskip

It is important to note that in an expansion in negative powers
of $a$, the $\phi$-dependent coefficient functions cannot in 
general be expected to display regular behaviour. This may be 
illustrated by expanding the function $(1+a^4 \phi^4)^{1/2}$
(which is $C^\infty$ on the domain $a>0$, $\phi={\rm arbitrary}$)
as $a^2\phi^2+ \frac{1}{2}a^{-2}\phi^{-2}+\dots$. For fixed $\phi$, the 
domain of convergence of this series 
is $a>|\phi|^{-1}$, which breaks down at $\phi=0$. 
The reason for limited convergence is that the function has
singularities at $a=\pm \sqrt{\pm i}\phi^{-1}$, against which a
condition like $C^\infty$ is insensitive. 
\medskip

However, we are free to {\it impose} analyticity properties
on the coefficient functions in the ansatz for $\psi(a,\phi)$, 
as long as they are compatible with the Wheeler-DeWitt equation. 
By doing so, we touch upon a deeper structure which is 
not yet completely understood. We choose as a first condition
the most natural requirement:
\par
{\it (i) The functions $F_r(\phi)$ and $G_r(\phi)$ are real
analytic, i.e. they are real and analytic in a neighbourhood of the
real axis in the compex $\phi$-plane. As a consequence, they admit
a Taylor expansion at $\phi=0$.} 
\par
\noindent 
Hence, by truncating the series in (\ref{generalansatz}) at
fixed $\phi$ and sufficiently large $a$ one should obtain a 
good numerical approximation for $\psi$. There will be a function
$R(\phi)$ such that the series converges for all $a>R(\phi)$.
This defines a domain of convergence in minisuperspace.
In order to exclude catastrophic behaviour of the
leading $a^3$ term in the exponent, we require in addition
\par
{\it (ii) $F_3(\phi)\le 0$ for all $\phi$.} 
\par
\noindent
The significance of this condition may be illustrated for the 
case of the approximate expressions for $\Omega_n^\pm$.
It allows for the dominant term 
$e^{-\xi^2/2}\equiv e^{- m a^3\phi^2/2}$ which appears
in $\Psi_n(\xi)$, 
whereas it prevents a behaviour of the type
$e^{\xi^2/2}\equiv e^{ m a^3\phi^2/2}$ which would otherwise
be possible. 
\medskip

Remarkably, the ansatz (\ref{generalansatz}) together with the
two conditions {\it (i)} and {\it (ii)} seem to leave only a
discrete freedom. We believe that the general solution is 
characterized by a non-negative integer $n$ and a choice of sign, 
i.e. a two-valued label $\pm$. The wave functions appearing
in this way are reckognized as exactifications of 
$\Omega_n^\pm(a,\phi)$ (and henceforth called by the same name)
that are distingushed by their analyticity properties. 
\medskip

Let us look at the sequence of equations generated by inserting
(\ref{generalansatz}) into (\ref{wdwposrep}) and dividing
by (\ref{generalansatz}). At the leading order $O(a^4)$ we
find
\begin{equation}
F_3'(\phi)^2 = 9 F_3(\phi)^2 + m^2 \phi^2\, . 
\label{Oa4}
\end{equation}
Condition {\it (ii)} specifies a unique solution. Let us suppose
that $F_3(0)\neq 0$. As a consequence, $F_3'(\phi)$,
which is given by a square root whose argument never vanishes,
will always be non-zero. For $\phi\rightarrow\pm\infty$ it will
either tend to $\infty$ or $-\infty$. In both cases it is not
possible for $F_3(\phi)$ to be non-positive for all $\phi$. 
Hence, we must have $F_3(0)=0$. 
This fixes $F_3'(0)=0$.
Further differentiation of (\ref{Oa4}) generates the
expansion 
\begin{equation}
F_3(\phi)=-m\left(
\frac{1}{2}\,\phi^2 + \frac{9}{32}\,\phi^4 + 
\frac{27}{256}\,\phi^6 + \dots\right)\, , 
\label{F3series}
\end{equation}
condition {\it (ii)} fixing the sign. This also agrees to leading
order with the behaviour of the exponential factor 
$e^{-\xi^2/2}\equiv e^{- m a^3\phi^2/2}$ in $\Psi_n(\xi)$. 
The behaviour of $F_3(\phi)$ for large $\phi$ is
\begin{equation}
F_3(\phi)\approx - 0.047\,m\, e^{3\phi} \, , 
\label{asexpF3}
\end{equation}
where the constant has been determined by numerical methods. 
\medskip

At the next order $O(a^3)$ we find the equation 
\begin{equation}
F_2'(\phi)F_3'(\phi) = 
6 F_2(\phi) F_3(\phi)\, . 
\label{Oa3}
\end{equation}
Since $F_3(\phi)$ is already uniquely determined,
$F_2(\phi)$ is fixed up to a multiplicative constant. 
In particular, we find $F_2'(0)=0$. 
The equation at $O(a^2)$, reads 
\begin{equation}
2 F_1'(\phi)F_3'(\phi)+F_2'(\phi)^2 = 
6 F_1(\phi) F_3(\phi) + 4 F_2(\phi)^2 - 1 \, . 
\label{Oa2}
\end{equation}
Again, $F_1(\phi)$ is determined only up to an
integration constant. 
Inserting $\phi=0$, we get $4F_2(0)^2=1$. Thus, there are
two possibilities $F_2(0) = \pm\frac{1}{2}$. 
For either sign, the solution of (\ref{Oa3})
is now uniquely determined to be
\begin{equation}
F_2(\phi)=\pm\left(
\frac{1}{2}+\frac{3}{4}\,\phi^2 +
\frac{45}{128}\,\phi^4+\frac{9}{512}\,\phi^6 + 
\dots\right)\, . 
\label{F2series}
\end{equation}
The leading behaviour of $F_2(\phi)$ for small $\phi$ thus 
reproduces $e^{\pm a^2/2}$, which we already know from (\ref{gsol2}). 
For large $\phi$ we find 
\begin{equation}
F_2(\phi)\approx \pm 0.231\, e^{2\phi} 
\label{asexpF2}
\end{equation}
as the leading asymptotic behaviour. 
\medskip

At $O(a)$, the number $q$ as well as the operator ordering
parameter $p$ and the first function of
the series in (\ref{generalansatz}) come into play. The
equation reads 
\begin{equation}
F_3''(\phi) + 2 F_3'(\phi)\,
\frac{G_0'(\phi)}{G_0(\phi)} + 
2 F_1'(\phi) F_2'(\phi) = 
4 F_1(\phi) F_2(\phi) + 3(2+p+2q) F_3(\phi)\, . 
\label{Oa1}
\end{equation}
Since, by condition {\it (i)}, $G_0(\phi)$ is analytic,
its Taylor expansion at $\phi=0$ exists and starts with 
$k_1 \phi^n + k_2\phi^{n+1}+\dots$, where $n$ is some
non-negative integer and $k_1\neq 0$. In the limit 
$\phi\rightarrow 0$ we have 
$F_3'(\phi)G_0'(\phi)/G_0(\phi)\rightarrow -m n$,
and (\ref{Oa1}) reduces in this limit to
$F_1(0)=\mp(n+\frac{1}{2})m\equiv \mp E_n$. As a consequence,
equation (\ref{Oa2}) admits a unique solution 
\begin{eqnarray}
F_1(\phi)=&\mp& E_n\, +\, \frac{3}{16 m}\,
\Big(-1\,\mp\, 4m E_n\Big)\phi^2\,
+\,\frac{9}{256m}\,\Big(3\,\mp\, 2m E_n\Big)\phi^4 +
\label{F1series1}\nonumber\\
& &\frac{3}{4096 m}\,\Big(41\,\mp\, 48 m E_n\Big)\phi^6 + \dots 
\label{F1series2}
\end{eqnarray}
The leading term thus reproduces $e^{\mp E_n a}$ from (\ref{gsol2}). 
For large $\phi$ we find 
\begin{equation}
F_1(\phi)\approx \mp 0.654\,E_n \, e^{\phi} 
\label{asexpF1}
\end{equation}
as the dominant behaviour. 
\medskip

From now on we just describe the general structure of the
subsequent steps. The limit $\phi\rightarrow 0$ of the
$O(a^0)$ equation yields that $\phi^{-n}G_1(\phi)$ must
be regular. Moreover, the leading order part of this
equation fixes
\begin{equation}
q = \frac{3n}{2}+\frac{1}{4}-\,\frac{p}{2} 
\mp\frac{1}{2}\, E_n^2\, . 
\label{q}
\end{equation}
The first term $3n/2$ combines together with $\phi^n$ from
$G_0(\phi)$ and a $m^{n/2}$ from the overall normalization 
into $\xi^n$ --- cf. (\ref{xi}) ---, which provides the leading 
order of the $n$-th Hermite polynomial $H_n(\xi)$. 
Taking into account the factor $a^{3/4}$ between the
position and oscillator representation (see (\ref{psihat})),
we exactly reproduce the contribution 
$a^{-p/2-1/2\,\mp \,E_n^2/2}$ from 
(\ref{omegaplusexp})--(\ref{omegaminusexp}). 
This result is inserted into (\ref{Oa1}), by which
$G_0(\phi)$ becomes unique up to the multiplicative constant 
$k_1$ which survives as an overall normalization
freedom for $\psi(a,\phi)$, and we find 
\begin{equation}
G_0(\phi)=k_1 \,\phi^n \left( 1 + 
\frac{3}{16m^2}\,\Big(3 m E_n-m^2\mp  1 \mp 2 m^2 E_n^2\Big)\phi^2+ 
\dots\right)\, . 
\label{G0series}
\end{equation}
The overall pattern seems to persist at all orders. 
All equations are of the linear inhomogeneous type,
leaving an integration constant which is determined at the
next order by the analyticity requirement.
We have checked this up to $G_7(\phi)$. 
\medskip

By re-writing the series in (\ref{generalansatz}) as
an exponential, one may re-arrange terms in a more
explicit way. This is in fact what 
Hawking and Page 
\cite{HawkingPage2} 
have done for the case $p=1$. When translated to our
formulation, their result seems to make explicit 
how the pattern determining the functions $G_r(\phi)$
persists to all orders. The uniqueness of the
coefficient functions is not considered as an important
issue in Ref.     
\cite{HawkingPage2}, 
but it is effectively achieved by throwing away 
$\ln(\phi)$-terms at each order). 
Since these authors intended to construct wormhole 
solutions, they considered only the exponentially
decreasing behaviour $e^{-a^2/2}$. 
They arrive at a set of functions $\Psi_n(a,\phi)$, which
we will denote as $\Psi_n^{\rm HP}(a,\phi)$. 
\medskip

Hence, without having a rigorous proof, we {\em conjecture} 
that all $G_r(\phi)$ exist and are uniquely determined, 
once the $\pm$ label and $n$ have been chosen. 
The case of the exponentially decreasing sector 
(lower sign) for $p=1$ seems to be covered by Ref. 
\cite{HawkingPage2}. 
Moreover, recalling the physical discussion of the approximate
wave functions in Section 3, we believe that
the series (\ref{generalansatz}) defines exact solutions
$\Omega_n^\pm(a,\phi)$ of the Wheeler-DeWitt equation for
all $(a,\phi)$, i.e. also outside the domain of convergence
of the series (in case this domain does not agree with
the whole of minisuperspace). 
We also note that, pulling an overall
factor $\phi^n$ out of $\Omega_n^\pm(a,\phi)$, only even powers
of $\phi$ remain, hence 
$\Omega_n^\pm(a,-\phi) = (-)^n\, \Omega_n^\pm(a,\phi)$. 
The functions 
$\Psi_n^{\rm HP}(a,\phi)$ as displayed by Hawking and Page
appearently coincide (up to normalization) with our 
$\Omega_n^-(a,\phi)$. 
(The leading order of their equation (72) is 
just the explicit formula 
for the $n$-th Hermite polynomial). 
\medskip

Due to the expansions of the coefficient functions
around $\phi=0$ we have an idea how 
$\Omega_n^\pm(a,\phi)$ behaves for
small $\phi$ and large $a$. By looking at the expression 
(\ref{asexpF3}) of $F_3(\phi)$ for large $\phi$, we
expect that, for sufficiently large and fixed $a$,
the dominant behaviour for large $\phi$ is
$\exp(- 0.047\,m a^3 e^{3\phi})$. If this conclusion holds,
the wave functions are actually more dampted than one would
expect from the factor 
$e^{-\xi^2/2}\equiv e^{-m a^3 \phi^2/2}$ of the oscillator
basis (\ref{hermite}) alone. 
One may in fact perform an independent 
analysis of solutions of the Wheeler-DeWitt equation 
in terms of large $\phi$, thus identifying the
$\Omega_n^\pm(a,\phi)$ by means of the asymptotic expressions 
(\ref{asexpF3}), (\ref{asexpF2}) and (\ref{asexpF1}). 
We will not go into these details but just 
complete our conjecture by noting that our wave
functions are likely to be well-behaved as 
$|\phi|\rightarrow\infty$. 
\medskip

\section{Exact solutions in the oscillator and
Fock representations}
\setcounter{equation}{0}

So far we have considered an expansion in $a^{-1}$ at constant
$\phi$. For actual computations the oscillator and
Fock representations turn out to be more suitable. 
As we have seen, the series in (\ref{generalansatz})
contains an overall factor $\phi^n$ which, together with 
$a^{3n/2}$ from (\ref{q}) 
makes up an overall factor $\xi^n$. Since the remainder
contains only even powers of $\phi$, the transformation
to the variable $\xi$ 
--- cf. (\ref{xi}) --- amounts to substitute 
$\phi^2\rightarrow m^{-1}a^{-3}\xi^2$, and even powers thereof. 
As a consequence, we never pick up half-integer powers of $a$ 
or negative powers of $\xi$.
Moreover, keeping $\xi$ fixed means
to follow a curve in minisuperspace whose $\phi$-coordinate
value decreases towards the axis $\phi=0$ as 
$a\rightarrow\infty$. This should not affect 
questions of existence and convergence very much
(or even improve the situation). 
The finite sum
$F_3(\phi)a^3+F_2(\phi)a^2+F_1(\phi)a$ reduces to
the expression $\pm a^2/2\pm E_n a$ plus a series of 
functions that contains only negative integer powers of $a$. 
\medskip

We can thus re-arrange the expression (\ref{generalansatz})
in terms of $a$ and $\xi$.
In order not to overcomplicate things, we assume the choice
of the $\pm$ sector and of $n$ has already been made, and
explicitly insert the leading orders. Transforming 
$\psi(a,\phi)$ into the oscillator representation 
$\widehat{\psi}(a,\xi)$ by (\ref{psihat}), we end up with
\begin{equation}
\widehat{\Omega}^\pm_n(a,\xi) = 
C^\pm_n \,a^{q_n^\pm} \,
\exp\left(\pm \frac{1}{2}a^2\mp E_n a\right)
\sum_{r=0}^\infty \frac{F_{rn}^\pm(\xi)}{a^r} \, , 
\label{omegaseries1}
\end{equation}
where the functions $F_{rn}^\pm(\xi)$ are analytic
and $C_n^\pm$ is an arbitrary overall normalization constant. 
The numbers $q_n$ may either be infered from (\ref{q}) by
taking into account the correct transformation factors, 
yielding 
\begin{equation}
q_n^\pm = -\,\frac{1}{2}-\,\frac{p}{2} \mp\frac{1}{2}\, E_n^2\, , 
\label{qn}
\end{equation}
or left undetermined in order to be re-discovered in 
the oscillator or Fock formalism. 
In view of our conjecture, the $F_{rn}^\pm(\xi)$ should be
uniquely determined. Note however that these
functions arise from a rearrangement of orders 
in the series of (\ref{generalansatz}), combined with
a series stemming from the exponent. Computationally, 
they are related to $F_r(\phi)$ and
$G_r(\phi)$ in a non-trivial way. 
Due to the discussion given in Section 3, 
we expect the eigenfunctions of $E$ to appear
at leading order in $a^{-1}$. Hence, we set 
\begin{equation}
F_{0n}^\pm(\xi) = \sqrt{n!}\, \Psi_n(\xi) \, , 
\label{F0n}
\end{equation}
which will be justified simply by being consistent. 
The prefactor $\sqrt{n!}$ is for later convenience. 
\medskip

This formulation is still a bit awkward for general $n$. 
We just report briefly on the expansion for $n=0$. 
Redefining the sum $\sum_r F_{r0}^\pm(\xi)/a^r$ as
$\exp(\sum_r g_r^\pm(\xi)/a^r)$, one finds that all
$g_r^\pm(\xi)$ are polynomials in $\xi$ containing only even
powers and being of order $\frac{2}{3}(r+3-j)$ with
$j=0$, $1$ or $2$. The formal criterion necessary to 
single out this unique sequence of functions turns out
to be the exclusion of homogeneous solutions for the
$g_r^\pm(\xi)$ of the error function type.
This just prevents a behaviour 
in the wave function that would contradict
condition {\it (ii)} of Section 4. 
\medskip

Some simplification occurs by translating the above 
expression (\ref{omegaseries1}) into the 
Fock representation in which states are written
as ${\cal F}(a,{\cal A}^\dagger)|0\rangle$, and this
is the setup we will consider now in more detail. 
The function (\ref{F0n}) is just $({\cal A}^\dagger)^n|0\rangle$.
All other $F_{rn}^\pm(\xi)$ --- which are of the form 
${\cal G}_{rn}^\pm({\cal A}^\dagger)|0\rangle$ --- are written as
$({\cal A}^\dagger)^n \,G_{rn}^\pm({\cal A}^\dagger)|0\rangle$, 
thereby defining a set of functions $G_{rn}^\pm({\cal A}^\dagger)$.
The wave functions thus become 
\begin{equation}
|\Omega_n^\pm,a\rangle = 
C^\pm_n \,a^{q_n^\pm}\, 
\exp\left(\pm \frac{1}{2}a^2\mp E_n a\right) ({\cal A}^\dagger)^n 
\sum_{r=0}^\infty \frac{G_{rn}^\pm({\cal A}^\dagger)}{a^r} 
|0\rangle \, , 
\label{omegaseries2}
\end{equation}
with 
\begin{equation}
G_{0n}^\pm({\cal A}^\dagger)\equiv 1\, . 
\label{G0n}
\end{equation}
According to the 
structure exhibited so far, we expect only even powers
of ${\cal A}^\dagger$ to occur in $G_{rn}^\pm({\cal A}^\dagger)$.
This is in accordance with ${\cal K}$ and
${\cal K}^2$ from (\ref{K2})--(\ref{Ksquare}), as appearing in
the Wheeler-DeWitt operator (\ref{wdwoperator2}), being even
in ${\cal A}$ and ${\cal A}^\dagger$.
\medskip

Due to our construction, the combinations 
$({\cal A}^\dagger)^n\,G_{rn}^\pm({\cal A}^\dagger)$
can be expected to be analytic at ${\cal A}^\dagger=0$.
One may however ignore the reasoning of Section 4 and
treat (\ref{omegaseries2}) and (\ref{G0n}) as an {\em ansatz
by its own} (leaving the numbers $q_n^\pm$ and $E_n$ unspecified 
as well). 
This is the strategy we will pursue in what follows. 
The procecure is again to separate orders of $a$ and to
determine the solutions by some 
additional requirement. One might impose analyticity of
$({\cal A}^\dagger)^n\,G_{rn}^\pm({\cal A}^\dagger)$, but it
turns out that a weaker condition does the job as well. 
We simply demand that 
\par
{\it (iii) $G_{rn}^\pm({\cal A}^\dagger)$ admits a
Laurent series at ${\cal A}^\dagger=0$.}
\par
\noindent
In other words, $G_{rn}^\pm({\cal A}^\dagger)$ may be expanded
in integer (positive and negative) powers of ${\cal A}^\dagger$. 
Logically, this replaces the condition {\it (i)}
as used in Section 4 (whereas an analogue of condition {\it (ii)} is
no longer necessary). 
The technical point of condition {\it (iii)} will be to 
exclude terms of the type $\ln({\cal A}^\dagger)$. 
\medskip

The Wheeler-DeWitt equation in the Fock representation is of 
fourth order in derivatives with respect to ${\cal A}^\dagger$. 
Although one might expect that this feature provides an
additional complication, things actually become simpler.
Writing a particular $|\Omega_n^\pm,a\rangle$ as
${\cal F}(a,{\cal A}^\dagger)|0\rangle$, we apply the
Wheeler-DeWitt operator and thereafter divide the
result by ${\cal F}(a,{\cal A}^\dagger)$. 
This allows for a proper separation of orders of $a$. 
Proceeding iteratively, one encounters only 
first order differential equations of a very simple 
type. Also, the solutions $G_{rn}^\pm({\cal A}^\dagger)$ 
turn out to be polynomials in positive and negative 
powers of $({\cal A}^\dagger)^2$, hence are represented in terms
of elementary functions. In other words, the Laurent series
whose existence is required by condition
{\it (iii)} are actually {\em finite}. This is a
great simplification as compared to the procedure of
Section 4, where the coefficient functions $F_r(\phi)$ and 
$G_r(\phi)$ emerged as infinite series. 
\medskip

The first non-trivial order $O(a)$ is of purely algebraic type
and yields $E_n=(n+\frac{1}{2})m$, thus re-introducting the
well-known eigenvalues of $E$. At $O(a^0)$ we encounter the 
differential equation 
\begin{equation}
2 m {\cal A}^\dagger \, 
\frac{d}{d {\cal A}^\dagger} 
G_{1n}^\pm({\cal A}^\dagger) =
\mp\,\frac{3n(n-1)}{2 ({\cal A}^\dagger)^2 } 
-E_n^2 \mp (1 + p + 2 q_n^\pm) 
\pm \frac{3}{2} ({\cal A}^\dagger)^2\, , 
\label{order0}
\end{equation}
the general solution of which is
\begin{equation}
G_{1n}^\pm({\cal A}^\dagger) = 
\pm \,\frac{3n(n-1)}{8m({\cal A}^\dagger)^2} 
+ \kappa_1
-\,\frac{1}{2m}\Big(E_n^2\pm(1+p+2 q_n^\pm)\Big) 
\ln({\cal A}^\dagger)
\pm \frac{3}{8m} ({\cal A}^\dagger)^2\, , 
\label{solutionG1n}
\end{equation}
where $\kappa_1$ is an arbitrary integration constant. 
Condition {\it (iii)} implies that the coefficient
of the $\ln({\cal A}^\dagger)$ term must vanish. This is an
equation for $q_n^\pm$, the solution immediately turning out 
to be (\ref{qn}). Inserting all results obtained so far 
into the equation at $O(a^{-1})$, we obtain
a differential equation of the type
\begin{equation}
{\cal A}^\dagger \, g'({\cal A}^\dagger) = 
\rho({\cal A}^\dagger)\, ,  
\label{type}
\end{equation}
where $g({\cal A}^\dagger)$ stands for 
$G_{2n}({\cal A}^\dagger)$
and $\rho({\cal A}^\dagger)$ for an expression that
has already been determined 
(up to the constant $\kappa_1$). Moreover, 
$\rho({\cal A}^\dagger)$ contains only integer powers,
ranging from $({\cal A}^\dagger)^{-4}$ to 
$({\cal A}^\dagger)^4$. The solution is thus a function of
equal type, including an additive integration
constant $\kappa_2$, and a $\ln({\cal A}^\dagger)$ term whose
coefficient turns out to be
\begin{equation}
\frac{1}{8 m^3}\left(-9 E_n^2-4 m^2 E_n^3 \mp 4 m^2 E_n
\pm 8 \kappa_1 m^2\right) \, . 
\label{coefflna}
\end{equation}
Again, this term has to vanish. Thus $\kappa_1$ is fixed,
and the complete solution for $G_{1n}^\pm({\cal A}^\dagger)$
is 
\begin{equation}
G_{1n}^\pm({\cal A}^\dagger) = 
\pm\, \frac{3n(n-1)}{8m ({\cal A}^\dagger)^2} 
+ \frac{E_n}{8m^2}\left(4m^2 \pm 9 \pm 4 m^2 E_n^2\right) 
\pm \frac{3}{8m} ({\cal A}^\dagger)^2\, . 
\label{solutionG1ntotal}
\end{equation}
This pattern persists at all orders. 
At $O(a^{-2})$ the differential equation for
$G_{3n}^\pm({\cal A}^\dagger)$ 
is again of the type (\ref{type}),
with $\rho({\cal A}^\dagger)$ being a finite sum of even integer 
orders of ${\cal A}^\dagger$, and involving the constant $\kappa_2$.
The solution for $G_{3n}^\pm({\cal A}^\dagger)$ thus consists of 
a finite sum of even integer orders of ${\cal A}^\dagger$, an
additive integration constant $\kappa_3$ and a
$\ln({\cal A}^\dagger)$ term whose coefficient has
to vanish, thus fixing $\kappa_2$, and so forth. 
We just display the highest and lowest order of the second 
coefficient function 
\begin{equation}
G_{2n}^\pm({\cal A}^\dagger) = 
\frac{9n(n-1)(n-2)(n-3)}{128 m^2({\cal A}^\dagger)^4} + 
\dots + \frac{9}{128 m^2} ({\cal A}^\dagger)^4 
\label{solutionG2ntotal}
\end{equation}
and note that in general $G_{rn}^\pm({\cal A}^\dagger)$ contains
contributions from $({\cal A}^\dagger)^{-2r}$ to
$({\cal A}^\dagger)^{2r}$. Moreover, the structure is such that
the combination 
${\cal G}_{rn}^\pm({\cal A}^\dagger) \equiv
({\cal A}^\dagger)^n\,G_{rn}^\pm({\cal A}^\dagger)$
is a polynomial. This is reflected by the coefficients
$n(n-1)$ and $n(n-1)(n-2)(n-3)$ in 
(\ref{solutionG1ntotal})--(\ref{solutionG2ntotal}). 
Thus, despite the expansion in powers of $a^{-1}$, 
the regular nature of the functional dependence
on ${\cal A}^\dagger$ remains intact. As already mentioned
above, such a feature is not at all generic for functions
that are regular in two variables, but it serves
here as part of the property singling out the wave functions 
$\Omega_n^\pm\,$. 
\medskip

Since the oscillatory domain is bounded in $a$ (and, moreover,
exists only if $n\gg m^{-2}$), there is no
analogous expansion there. However, due to the WKB matching
procedure as applied in Section 3, we {\em define} another
set of {\em exact} wave functions $\Xi_n^\pm$ by (\ref{Xidef})
(or, inversely by (\ref{OmXi1})--(\ref{OmXi2})), where
$\Theta_n$ is still given by (\ref{Theta}). In order to
achieve the (approximate) oscillatory behaviour as in
(\ref{Xiosc}), we have to define the normalization
constants in (\ref{omegaseries1}) and (\ref{omegaseries2}) 
so as to give $\Omega_n^\pm$ the asymptotic
form (\ref{omegaplusexp})--(\ref{omegaminusexp}), hence 
\begin{equation}
C_n^+ = \frac{1}{\sqrt{n!}} 
\left(\frac{2}{E_n}\right)^{-E_n^2/2} e^{E_n^2/4}
\qquad\qquad 
C_n^- = \frac{1}{2\sqrt{n!}} 
\left(\frac{2}{E_n}\right)^{E_n^2/2} e^{-E_n^2/4} \, . 
\label{normconst}
\end{equation}
This completely determines our wave functions $\Omega_n^\pm$ 
and $\Xi_n^\pm\,$. Due to their structure 
(the appearance of the oscillator
{\em basis} $|n\rangle$ at leading order) it is clear that
a rather large set of solutions of the Wheeler-DeWitt
equation may be expanded into them. By assuming
\begin{equation}
\psi = \sum_{n=0}^\infty 
( k^+_n\, \Omega^+_n  + k^-_n\, \Omega^-_n )
= \sum_{n=0}^\infty 
( c^+_n\, \Xi^+_n  + c^-_n\, \Xi^-_n )
\label{expansioninXi}
\end{equation}  
for some solution $\psi$ of the Wheeler-DeWitt equation, 
the $k_n^\pm$ (or $c_n^\pm$) may be computed by fixing some 
(large enough) $a$
and expanding the initial data $\psi$ and $\partial_a\psi$
in terms of the $\Omega_n^\pm$ {\em at} the fixed
value of $a$ 
(which should be possible if the bevaviour in $\phi$ is
not too catastrophic).  
We will find a more convenient method later on, and the
present argument just serves for the count of degrees of freedom
contained in $\Omega_n^\pm\,$. Thus we treat $\{\Omega_n^\pm\}$
(or likewise $\{\Xi_n^\pm\}$) as a {\em basis} of the
space $\H$ of wave functions, and a precise definition of which
coefficients in (\ref{expansioninXi}) are
allowed will be given later. 
The $\Omega_n^\pm$ are, by construction, real 
(complex conjugation ${}^*$ being defined by its action in the 
position representation), while the transformation of the 
basis (\ref{Xidef}) implies
\begin{equation}
(\Xi_n^\pm)^* = \Xi_n^\mp \, . 
\label{reality}
\end{equation}
If $n\gg m^{-2}$ and at the level of the WKB approximation
at which (\ref{Xiosc}) is valid, this property is 
in accordance with $\Xi_n^+$ and $\Xi_n^-$
representing an ensemble of collapsing and expanding 
universes, respectively. 
\medskip 

If one accepts the wave functions $\Xi_n^\pm$ to play a 
distunguished role, one ends up with a distinguished decomposition
of the space $\H$ of wave functions into the span 
$\H^+$ of $\{\Xi_n^+\}$ and the span
$\H^-$ of $\{\Xi_n^-\}$, hence $\H=\H^+\oplus\H^-$. 
Wave functions are thus uniquely decomposed as
$\psi=\psi^+ + \psi^-$. 
It is important here to note that the differential structure
of minisuperspace is not sufficient for
identifying collapsing and expanding modes exactly.
This is a major difference to the flat 
Klein Gordon equation (where a differential background 
structure, namely a timelike Killing vector field, enables one
to define negative/positive frequency modes exactly
in a Lorentz-invariant way) and it provides one of the
most severe problems in constructing a consistent
setup for quantum cosmology. 
However, if the {\em analyticity structure} of wave functions
is accepted as a guiding principle in our model, we seem
to have uniquely defined a decomposition of the
space of wave functions which 
--- in the WKB approximation ----
is identified with collapsing and expanding (incoming
and outgoing) modes.
\medskip

There is a formal relation between $\Omega_n^+$ and
$\Omega_n^-$ that might provide a hint towards a deeper 
significance of the analytic structure we have exhibited so far. 
The Wheeler-DeWitt equation remains invariant under 
the substitution
\begin{equation}
a\rightarrow i a \qquad\qquad
m\rightarrow i m\, , 
\label{substitution}
\end{equation}
leaving $\phi$ unchanged. Thus the quantities $m a^3$
and $m^{-1} a$ (and by definition $\xi$, which involves
$m^{1/2} a^{3/2}$)
remain unchanged as well, but we have $a^2\rightarrow - a^2$,
and $m^2\rightarrow - m^2$. Under this substitution the
wave functions $\Omega_n^+$ and $\Omega_n^-$, when written
down in the large-$a$ expansion, are almost
perfectly transformed into each other. This includes the
structure of the exponential prefactors as well as the
normalization (\ref{normconst})
(cf. (\ref{omegaplusexp})--(\ref{omegaminusexp}) and
note that $E_n^2\rightarrow-E_n^2$ while
$a/E_n\rightarrow a/E_n$), 
except for the prefactor $a^{-1/2-p/2}$ and the numerical factor
$\frac{1}{2}$ in $\Omega_n^-$. 
(Also Hawking and Page have noted that this transformation
carries their $\Psi_0^{\rm HP}(a,\phi)$ --- which is
our $\Omega_0^-(a,\phi)$ --- into the exponentially
growing part of the no-boundary wave function as given
by Page 
\cite{Page} 
--- which is just our $\Omega_0^+(a,\phi)$). 
A similar relation exists 
between the Airy functions 
${\rm Ai}(x)$ and ${\rm Bi}(x)$ in the expansion for
large $x$, if $x\equiv a^{4/3}$ is set. 
One could thus try to define $\H^\pm$ as the
''eigenspaces'' under a suitable substitution operation. 
However, this forces us to treat $m$ as a variable 
rather than a fixed constant. We leave it open whether 
one would gain any new insight by doing so. 
\medskip

The crucial question in exploiting the emergence of
the spaces $\H^\pm$ for further developments of the subject 
of quantum cosmology is certainly whether the structure showing 
up here carries over to the full (non-minisuperspace)
Wheeler-DeWitt equation. 
\medskip

\section{Scalar product and Hilbert spaces}
\setcounter{equation}{0}

There is a natural (Klein Gordon type) scalar product 
associated with the Wheeler-DeWitt equation
\cite{DeWitt}\cite{FE8}. 
If $\psi_1$ and $\psi_2$ are two solutions of the
latter that are well-behaved for large $\phi$, the
expression 
\begin{equation}
Q(\psi_1,\psi_2) = -\,\frac{i}{2} \,a^p
\, \int_{-\infty}^\infty
d\phi\, \left(
\psi_1^*(a,\phi)
\stackrel{\leftrightarrow}{\partial_a} \psi_2(a,\phi)
\right)
\label{Qhm}
\end{equation}  
(with $f \stackrel{\leftrightarrow}{\partial_a} g \equiv
f\,\partial_a  g - (\partial_a f)\,g)$
is independent of $a$ on account of (\ref{wdwposrep}). 
It defines an indefinite scalar product, the integrand being the 
$a$-component of a conserved current 
\cite{Vilenkininter}. 
Due to its indefiniteness, it does not enable us to define a
Hilbert space directly. 
\medskip

When wave functions are expressed in the energy representation 
as introduced in Section 2, we find
\begin{eqnarray}
Q(\psi_1,\psi_2)&=&
-\,\frac{i}{2} \,
a^p\, \sum_{n=0}^\infty
\Bigg(
f_n^*(a) \stackrel{\leftrightarrow}{\partial_a} g_n(a)+ 
\nonumber\\
& & \frac{3}{2a}\,\sqrt{n(n+1)}
\left( f_{n-1}^*(a)g_{n+1}(a) - f_{n+1}^*(a) g_{n-1}(a) 
\right) \Bigg) \, , 
\label{Qhmenergy}
\end{eqnarray}
where $\widehat{\psi}_1(a,\xi)=\sum_n f_n(a)\Psi_n(\xi)$
and $\widehat{\psi}_2(a,\xi)=\sum_n g_n(a)\Psi_n(\xi)$.
In the Fock representation the scalar product reads
\begin{equation}
Q(\psi_1,\psi_2) = -\,\frac{i}{2}\, 
a^p \,\langle\psi_1,a| \, 
\stackrel{\leftrightarrow}{\partial_a} + \, 
\frac{3}{a}\,{\cal K} \,\,|\psi_2,a\rangle\, , 
\label{QhmFock}
\end{equation}  
where the derivative $\stackrel{\leftrightarrow}{\partial_a}$,
when acting to the left, does not include the prefactor
$a^p$. 
\medskip

Inserting the asymptotic expressions of our wave functions
in the oscillatory domain, we find 
\begin{equation}
Q(\Omega^\pm_r,\Omega^\pm_s) = 0 \qquad\qquad
Q(\Omega^+_r,\Omega^-_s) = \delta_{rs}  
\label{basis1}
\end{equation}
and 
\begin{equation}
Q(\Xi^\pm_r,\Xi^\pm_s) = \pm \delta_{r s}
\qquad\qquad
Q(\Xi^+_r,\Xi^-_s) = 0\,.
\label{basis2}
\end{equation}
Using the asymptotic series for $a\rightarrow \infty$, we find
hints that these relations hold {\em exactly} for all
$r$ and $s$. For all those combinations in which the
exponential $a$-dependent prefactor decreases, they are
evident ($Q$ being evaluated at arbitrarily large $a$). 
For all other cases, we have used the first few terms
of the series to check them. In the following we will
assume that they hold. They also explain the
particular normalization we have choosen (see
(\ref{Xiosc}) and (\ref{normconst})). 
\medskip

Let us consider now the decomposition 
$\H=\H^+\oplus\H^-$ of the space of
solutions of the Wheeler-DeWitt equation as induced by
the wave functions $\Xi_n^\pm$. If any solution $\psi$ is
expanded as in (\ref{expansioninXi}), the above normalization
of the basis yields 
\begin{equation}
c_n^\pm = \pm Q(\Xi^\pm_n,\psi)\, . 
\label{expansioncoeff}
\end{equation}  
Encoding the information contained in a wave function
in terms of the numbers $c_n^\pm$, we can talk about
an ''energy representation'' in a sense quite more sophisticated 
than the notation of Section 2. Since the basis elements
are solutions of the Wheeler-DeWitt equation, there is
no dependence on additional variables, except for the
''true'' physical labels $n$ and $\pm$. 
\medskip

Evidently, the scalar product $Q$ is positive/negative
definite on $\H^+$ and $\H^-$, respectively. 
Hence, admitting only wave functions $\psi$ whose coefficients
satisfy
\begin{equation}
\sum_{n=0}^\infty (\,|c_n^+|^2+|c_n^-|^2\,) < \infty 
\label{coeffnorm}
\end{equation}
makes $(\H^+,Q)$ and $(\H^-,-Q)$ two Hilbert spaces that may
be used as a starting point for despriptions how to
compute probabilities for observations. 
Admitting more general wave functions, one should still
be able to compute {\it relative} probabilities. Given a
solution $\psi$ as in (\ref{expansioninXi}), the
(relative) probabilities associated with the states
$\Xi_n^\pm$ are $|Q(\Xi_n^\pm,\psi)|^2$. These numbers
should be relevant when predictions for matter the energy 
contents are drawn. In the WKB-philosophy --- which
is relevant for quantitative predictions --- this is 
quite clear for {\em any} basis being normalized
as (\ref{basis2}). (Any such basis gives rise to
a decomposition into two Hilbert spaces). 
In Ref. 
\cite{FE8} 
we have suggested a ''minimal'' interpretational scheme 
for quantum cosmology based
on $(\H,Q)$ as the {\it only} fundamental mathematically
well-defined structure. 
However, here we suggest a different thing. 
If one accepts the basis wave functions
$\Xi_n^\pm$ as distinuished objects (not just as
wave functions which describe a semiclassical ensemble of 
universes with approximately conserved matter energy), one 
is faced with a {\it distinguished Hilbert space structure}, based
on asymptotic analyticity properties. 
One might think of it to be introduced by some path-integral 
or to constitute a first principle by its own, and it
constitutes an exact fundamental structure
{\em in addition} to $(\H,Q)$. 
If an analogous feature 
may be found in the full (non-minisuperspace) case as well, 
it should be of some relevance to the fundamental conceptual 
problems of quantum cosmology. 
\medskip

\section{No-boundary wave function}
\setcounter{equation}{0}

In order to show an example, we estimate the
coefficients of the expansion (\ref{expansioninXi})
for the no-boundary wave function 
\cite{HartleHawking} 
$\psi_{\rm NB}$ of the Hawking model
\cite{Hawking2}.  
It has been studied in great detail by Page
\cite{Page}. 
The classical no-boundary trajectories start from a
point $(a_0,\phi_0)$ on the zero potential line
$m^2 a^2 \phi^2=1$ and are in the inflationary domain
$m a |\phi|\gg 1$, $|\phi|\gg 1$ 
given by
\begin{equation}
a = \frac{1}{2 m \phi_0^{2/3}\phi^{1/3}}
\, \exp\left(\frac{3}{2} ( \phi_0^2 - \phi^2 )\right)
\approx 
\exp\left(\frac{3}{2} ( \phi_0^2 - \phi^2 )\right).
\label{trajhm1}
\end{equation}
This expression is valid as long as 
$1\ll|\phi|\ll|\phi_0 - 1/(3 \phi_0)|$.
Since there is only one no-boundary trajectory hitting any
point $(a,\phi)$ in the inflationary domain, we associate
with this point the initial value $\phi_0$ of the latter,
thus turning $\phi_0$ into a function of $a$ and $\phi$. 
The action of this congruence of trajectories is 
\begin{equation}
S=  -\,\frac{1}{3 m^2\phi^2} \left( m^2 a^2 \phi^2-1\right)^{3/2}
+\, \frac{\pi}{4} 
\approx -\, \frac{1}{3} m |\phi| a^3\, , 
\label{action}
\end{equation}
the no-boundary wave function in the inflationary domain
being (thereby generalizing Page's expression to arbitrary $p$)
\begin{equation}
\psi_{\rm NB}(a,\phi) \approx a^{-p/2-1}\, A(\phi_0)\, 
 \cos(S(a,\phi))
\label{nobound}
\end{equation}
with 
\begin{equation}
A(\phi_0) = 
\frac{1}{ \sqrt{\pi m |\phi_0|} } 
\left(
   \sqrt{6} + 2 \,\Big( \exp(\frac{1}{3m^2\phi_0^2})-1 \Big) 
\right) 
\label{noboundprefac}
\end{equation}
and $\phi_0$ now interpreted as a function of $a$ and $\phi$. 
Its basic structure is that it is a product of the
{\it rapidly} oscillating WKB-type function $\cos(S)$ with 
a {\it slowly} varying prefactor. 
(The prefactor $A(\phi_0)$ is in fact to some extent
arbitrary. In the WKB-approximation the Wheeler-DeWitt
implies that it is constant along the classical 
trajectories. The expression (\ref{noboundprefac}) 
corresponds to the solution arising from the
no-boundary proposal for the Euclidean path-integral). 
\medskip

The no-boundary trajectory with initial value $\phi_0$ 
leaves the inflationary domain (i.e. attains
$|\phi|\approx 1$) at 
$a_{\rm min}\approx m^{-1}|\phi_0|^{-2/3} e^{3\phi_0^2/2}$,
and subsequently undergoes the matter dominated
era in which $\phi$ oscillates and the matter energy $E$ 
is approximately conserved. Since, on the other
hand, $a_{\rm min}\approx(n/m)^{1/3}$, we find 
$\phi_0^2\approx \frac{2}{9}\ln(n m^2)$, 
which makes $\phi_0$ in (\ref{nobound}) effectively a function
of $n$. Since $|\phi_0|\gg 1$, we have 
$n\gg m^{-2}$, and thus a non-trivial classical domain. 
The prefactor (\ref{noboundprefac}) then becomes 
\begin{equation}
A_n \approx 
\sqrt{\frac{3}{\pi m}}\,
\frac{1}{\sqrt[4]{2\ln(n m^2)}}
\left(
   \sqrt{6} + 2 \,\Big( \exp (
\frac{3}{2m^2\ln(n m^2)} )-1 \Big) 
\right) 
\label{prefacn}
\end{equation}
as far as the contribution of trajectories representing
universes with matter energy $E_n$ is concerned.
\medskip

We will choose an indirect way to estimate the magnitude
of the coefficients $c_n^\pm$ when $\psi_{\rm NB}$ is expanded
into $\Xi_n^\pm$ as in (\ref{expansioninXi}).
Since we do not know precisely the behaviour of 
$\psi_{\rm NB}(a,\phi)$ in just those regions in which we know
$\Xi_n^\pm(a,\phi)$, the information necessary to perform
the integration (\ref{expansioncoeff}) is not easily accessible. 
We may instead first evaluate the relative probability
distribution for trajectories labelled by $\phi_0$ in
the position representation. Since $Q(\psi,\psi)=0$ for
real $\psi$, we consider the incoming/outgoing projections
$\psi_{\rm NB}^\pm\approx\frac{1}{2}a^{-p/2-1} A e^{\mp i S}$.
The expressions $Q(\psi_{\rm NB}^\pm,\psi_{\rm NB}^\pm)$, when
computed according to (\ref{Qhm}), turn out to be
integrals over the measure 
$\pm\frac{1}{4} m A^2\, |\phi| \, d\phi$ (the additional
$a$-dependence cancelling, as it should).
Using $|\phi|\,d\phi\approx|\phi_0|d\phi_0$ (at constant $a$),
this becomes $\pm P(\phi_0)\,d\phi_0$
($\pm$ the relative probability for finding
the universe represented by a trajectory in the
interval between $\phi_0$ and $\phi_0+d\phi_0$), where
\begin{equation}
P(\phi_0)= \frac{m}{4}\,|\phi_0| \, A^2(\phi_0)\, . 
\label{measureP}
\end{equation}
This is the standard procedure of 
evaluating probabilities for 
WKB-type wave functions based on the conserved current.
(An alternative interpretation
\cite{HawkingPage} 
predicts probabilities that differ from these by the
amount of proper time spent by trajectories in a
domain of minisuperspace). 
By including an additional
factor 2, we may restrict $\phi_0$ to be positive. 
Since in this case $\phi_0$ and $n$ are related uniquely by
$\phi_0\approx \frac{1}{3}\sqrt{2 \ln(n m^2)}$, 
we may compute $d\phi_0/dn \approx (9n\phi_0)^{-1}$. 
Due to the symmetry 
$\psi_{\rm NB}(a,-\phi)=\psi_{\rm NB}(a,\phi)$
we know that $c_n^\pm=0$ for odd $n$. Supposing that
$|c_n^\pm|$ for even $n$ may be approximated by continuous 
functions, we set $dn=2$ and find that the 
relative probability for the universe to have
energy quantum number $n$ is given by 
$P_n\approx 4 \,(9n\phi_0)^{-1} P(\phi_0) $.
On the other hand, the expressions 
$Q(\psi_{\rm NB}^\pm,\psi_{\rm NB}^\pm)$ are given
by $\pm \sum_n |c_n|^\pm$, and the relative
probability to find the universe containing matter energy 
$E_n$ in the contracting/expanding mode is
$P_n^\pm=\frac{1}{2}P_n=|c_n^\pm|^2$ for even $n$ (and zero for
odd $n$). 
We thus identify for even $n$ and $n\gg m^{-2}$ 
\begin{eqnarray}
c_n^+ &=& (c_n^-)^* \approx 
\frac{1}{3}\,\sqrt{\frac{2 P(\phi_0)}{n\phi_0}}\,K_n 
\approx\frac{1}{3}\,\sqrt{\frac{m}{2n}}\, A_n\, K_n \approx 
\nonumber\\
& & 
\frac{1}{\sqrt{6\pi n}}\,
\frac{1}{\sqrt[4]{2\ln(n m^2)}}
\left(
   \sqrt{6} + 2 \,\Big( \exp (
\frac{3}{2m^2\ln(n m^2)} )-1 \Big)  
\right) K_n 
\label{cn}
\end{eqnarray}
where $|K_n|=1$. 
The first equality is exact, it stems from the fact that
$\psi_{\rm NB}$ is real. 
(Note that due to the reality of $\psi_{\rm NB}$,
both $\pm$ modes are of equal probability, although this 
number is infinite). 
\medskip

The density $P(\phi_0)$ as well as the
probabilities $P_n$ display the well-known problems
for the no-boundary wave function to predict sufficient 
inflation (i.e. a sufficienty large universe; see Refs. 
\cite{GrishchukRozhansky}\cite{Lukas}). 
Due to the smallness of $m$, the exponentials
prefer small $\phi_0$ and $n$, and it is only the flat 
behaviour of $P(\phi_0)$ for $\phi_0\rightarrow\infty$
(or the dominant behaviour $P_n\sim n^{-1}$ as 
$n\rightarrow\infty$) that allow for large universes.
However, trajectories with $m |\phi_0|\gsim 1$ correspond
to classical universes above Planckian densities after
nucleation, and it is not clear whether they should
contribute (cf. Ref. 
\cite{HawkingPage}). 
These trajectories correspond to
$n\gsim m^{-2} \exp(\frac{9}{2m^2})\approx\exp(5\times10^{12})$
(and thus $a_{\rm max}$ being given roughly by the same
number, or, expressed in ''true'' units as displayed in the 
Introduction, 
$\widetilde{a}_{\rm max}\approx \exp(5\times 10^{12})$ centimeters, 
or light years, or present Hubble scales,
which makes no big difference due to the  huge value of this 
number). 
\medskip

As a consequence of (\ref{cn}), we find 
in the energy representation
$|\psi_{\rm NB},a\rangle=\sum_n f_n(a)|n\rangle$ that
the oscillator components (for even $n\gg m^{-2}$ and in 
the range $a_{\rm min}\approx (n/m)^{1/3}\lsim a\ll
a_{\rm max}\approx 2 m n$) are given by 
\begin{equation}
f_n(a)\approx \frac{2}{3}\, a^{-p/2-1/4} \, 
\sqrt{\frac{m}{2n}}\, 
\frac{A_n}{\sqrt[4]{2 E_n}}\, 
\cos\left(
\frac{2}{3}\,(2E_n)^{1/2} a^{3/2} + \delta_n\right)\, . 
\label{fnanob}
\end{equation}
Here, we have set $K_n=e^{i \delta_n}$. 
Due to the large value of $n$ one may of course set
$E_n\approx m n$. 
The components for odd $n$ vanish. 
For $a\gsim a_{\rm max}$ the $f_n(a)$
contain the exponential $e^{\pm a^2/2\mp E_n a}$ terms 
of (\ref{omegaplusexp})--(\ref{omegaminusexp}). 
An analogous large-$a$ behaviour is expected 
to apply for the component functions with $n\lsim m^{-2}$ 
(which do not contribute to classical universes) as well. 
\medskip

The expression (\ref{fnanob}) may heutistically be checked 
by using the real part of (\ref{eikxi}) with
$k=\frac{1}{3}m^{1/2}a^{3/2}$ as an approximation for
$\cos(S)$. Invoking
$\Psi_n(x)\approx\Psi_n(0) \cos(x\sqrt{2n})$ for small $x$
and even $n$, 
and $\Psi_n(0)\approx (-)^{n/2} (2/n)^{1/4} \pi^{-1/2}$
for large even $n$ (which follows from Stirling's formula
for $n!$), 
a behaviour roughly similar to
(\ref{fnanob}) is recovered but without the phases $\delta_n$
and an additional factor $\frac{1}{2}$ in the Cosine. 
This reflects our lack of knowledge about the details 
of $\psi_{\rm NB}(a,\phi)$ in the domain where 
$\Xi_n^\pm(a,\phi)$ is known, and {\it vice versa}. 
It would be interesting to study this problem in more
detail, and to find an estimate for the $\delta_n$.
In case these numbers vary rapidly with $n$, the Cosine
in (\ref{fnanob}) would introduce a chaotic type behaviour
of the oscillator excitations.
\medskip

In their work on wormholes Hawking and Page
\cite{HawkingPage2} 
have assumed that the no-boundary wave function
(which increases as $e^{a^2/2}$)
can be expanded in terms of their wave functions
$\Psi_n^{\rm HP}(a,\phi)$, 
which are our $\Omega_n^-(a,\phi)$ and decrease as
$e^{-a^2/2}$. In our language, they have assumed 
$\{\Omega_n^-\}$ to form a basis, in which case one would
expect the expansion coefficients showing a
tremendous increase with $n$. 
In contrast, in our framework, these states are only
half of a basis. Note that linear independence relies
on the precise definition of a vector space. 
(For example, Sine and Cosine functions can or cannot be 
expanded into each other, depending on the interval on which
they are considered).
However, by just counting degrees of freedom, 
the existence of a framework in which 
$\psi_{\rm NB}$ may be expanded in terms of $\{\Omega_n^-\}$ 
is very unlikely. (This would in fact imply that
$\Omega_n^+$ may be expanded in terms of $\Omega_n^-$, 
and the structure provided by the scalar product $Q$ would
break down). The explicit computation of the expansion
coefficients $k_n^\pm$ with respect to $\Omega_n^\pm$ as defined by
(\ref{expansioninXi}) seems to be difficult at the present status of
our knowledge. For $c_n^+ = (c_n^-)^*$, we find 
$k_n^+ +i k_n^- = 2 c_n^+ e^{i\Theta_n}$ with
$\Theta_n$ from (\ref{Theta}). Our estimate
(\ref{cn}), together with $K_n=e^{i \delta_n}$, leads to 
a real expression multiplied by $e^{i(\Theta_n+\delta_n)}$.
Since we do not know $\delta_n$, we cannot compute the
real and imaginary part of this phase factor directly. On 
the other hand, an estimate for $\delta_n$ might be accessible
by using further information about $\psi_{\rm NB}$. The mere
fact that it containts an increasing $e^{a^2/2}$ contribution
implies that at least some $\Theta_n+\delta_n$ are different
from $\frac{\pi}{2}$ (modulo $2\pi$).
\medskip

There is another interesting question related with
the work of Hawking and Page. They assume that
the regular superpositions of $\{\Omega_n^-\}$ 
--- due to their exponentially dampted nature --- 
provide quantum wormhole states. 
On the other hand, the second half
of the basis $\{\Omega_n^+\}$ appears on quite an
equal footing here: both types of states describe universes with 
matter energy $E_n$. The only possible essential difference 
concerns observations when the universe is near its classical 
turning point $a\approx a_{\rm max}\,$. 
It is in particular the states $\Omega_n^-$ that 
do not seem to provide problems there. How does this fact
relate to the interpretation of certain superpositions
of these states as wormholes? 
Usually, the exponentially damped behaviour $e^{-a^2/2}$
is associated with wormhole states by definition. 
However, we do not have a satisfactory interpretation
of the exponentially increasing behaviour $e^{a^2/2}$
(nor is it required {\it by definition} for any
wave function, but just accepted as a grain of salt rather
than a desired property when it emerges). 
This is a certain asymmetry (at least as long as
no path-integral arguments are invoked) that might 
point towards 
a theoretical lack in our understanding of quantum 
cosmology. We are not able to give an
answer to this question, but it seems worth pursuing it. 
\medskip

The predictive power contained in the coefficients
(\ref{cn}) is --- as far as practical quantitative features
are concerned --- equal to the results of the common 
WKB-philosophy. If, however, the decomposition $\H=\H^+\oplus\H^-$ 
(with $\H$ being defined by the normalization condition
(\ref{coeffnorm})) is regarded as a distinguished one, we 
may embed $\psi_{\rm NB}$ into a mathematically well-defined 
underlying structure. We have {\em two fundamental Hilbert spaces}
$(\H^\pm,\pm Q)$ and a wave function for which
$Q(\Xi,\psi_{\rm NB})$ is finite for any element
$\Xi\in\H$, and it is just these numbers in
which all physical information about $\psi_{\rm NB}$
is encoded. The wave function $\psi_{\rm NB}$ is not an element 
of $\H$. Since both projections $\psi_{\rm NB}^\pm$ onto $\H^\pm$ 
have infinite norm $Q(\psi_{\rm NB}^\pm,\psi_{\rm NB}^\pm)$,
they are of distributional character (similar to momentum
eigenstates in conventional quantum mechanics). 
\medskip

\section{Concluding remarks}
\setcounter{equation}{0}

The behaviour of the real wave functions 
$\Omega_n^\pm(a,\phi)$ for large $a$ provides a
relation between the expansion 
(\ref{expansioninXi}) of some $\psi$ into these 
(the coefficients $k_n^\pm$
being trivially connected to $c_n^\pm$ by
(\ref{Xidef})--(\ref{OmXi2})) and the question of
boundedness of $\psi$.
Due to the factors $e^{a^2/2}$, 
all $\Omega_n^+(a,\phi)$ are unbounded. Hence,
a given wave function $\psi(a,\phi)$ seems to be
bounded away from $a=0$ (i.e. $|\psi(a,\phi)|<K<\infty$
in any domain $a>a_1$) if and only if 
$k_n^+=0$ for all $n$. 
(Note that this is in contradiction with the remark in Ref. 
\cite{HawkingPage2} 
concerning the expectation that $\psi_{\rm NB}$, which is not
bounded, may be expanded in terms of 
$\Psi_n^{\rm HP}\equiv \Omega_n^-$). 
It is not entirely clear to what extent
the unboundedness of a wave function causes interpretational 
problems (e.g. near the classical turning point). 
Some authors consider boundedness as a 
condition necessary for interpretation 
\cite{Zeh}\cite{Kiefer1} 
and talk about a ''final condition'' for the wave function. 
Such approaches could provide an additional justification 
for expecting the limit $a\rightarrow\infty$ to play a 
conceptually fundamental role. 
At the technical level we do not know whether the degrees of 
freedom contained in the coefficients $k_n^-$ may be arranged 
so as to cancel the expected singular behaviour of 
$\Omega_n^-(a,\phi)$
for small $a$ and make up a strictly bounded solution 
(although Hawking and Page, when constructing approximate
wormhole states, provide a hint that this is possible). 
In the case of the tunnelling wave function emerging from the
outgoing mode proposal
\cite{Lindeout}\cite{Vilenkintunnell}, 
the boundedness of $\psi_{\rm T}$ is usually
considered part of the definition. If such functions
exist at all, 
this would immediately imply that $\psi_{\rm T}$
is a superposition of the $\Omega_n^-$ alone. 
It might be worth thinking about whether a possible 
relation between $\psi_{\rm T}$ and the wormhole 
context in which Hawking and Page 
considered the wave functions 
$\Psi_n^{\rm HP}\equiv\Omega_n^-$ sheds some new
light on the conceptual questions. 
\medskip

Let us close this article by adding some general 
speculations. If the structure encountered in the Hawking 
model carries over to some more sophisticated 
(preferably non-minisuperspace) model one would apply 
WKB-techniques in combination with decoherence arguments 
(traces in the Hilbert spaces $\H^\pm$)
in order to identify states with physical
observables and to recover the standard laws of 
physics. The $Q$-product should boil down 
to plus or minus the standard scalar product of
quantum mechanics
\cite{DeWitt}\cite{FE8}.  
The existence of two separate Hilbert spaces may be a hint
that the $\pm$ sectors decouple from any observational
point of view: given a wave function, one is {\it either}
in the $+$ {\it or} in the $-$ sector, no experience of a
superposition is possible. One could call this a 
super-selection rule. Only {\it within} these sectors 
Hilbert space techniques apply, 
and the actual non-experience of various other superpositions
is delegated to decoherence. The ultimate
object to describe experience would be a reduced density matrix, 
as evaluated by standard Hilbert space methods, 
hence within a completely well-defined framework. 
It remains to be seen whether
such an interpretation is still possible when observations
near the turning point are concerned.
There, one would heuristically expect to undergo a
''transition'' from a reduced density matrix
belonging to the $-$ sector to one belonging to the
$+$ sector. 
\medskip

Since the situation is a bit reminiscent of the one-particle
Hilbert spaces of negative/positive frequency modes
as emerging from the flat Klein Gordon equation,
a further possible direction to pursue is to envisage a 
third-quantization 
\cite{HosoyaMorikawa}\cite{McGuigan}
in terms of the preferred decomposition
(cf. Ref. \cite{FE8}). 
\\
\\
\\


\begin{thebibliography}{00}

\bibitem{DeWitt}
{
 B. DeWitt,
 ''Quantum Theory of Gravity. I. The Canonical Theory'',
 {\it Phys. Rev.} {\bf 160}, 1113 (1967).
}

\bibitem{Hawking2}
{
 S. W. Hawking,
 ''The quantum state of the universe'',
 {\it Nucl. Phys. B} {\bf 239}, 257 (1984).
}

\bibitem{HawkingPage2}
{
 S. W. Hawking and D. N. Page,
 ''Spectrum of wormholes'',
 {\it Phys. Rev. D} {\bf 42}, 2655 (1990).
}

\bibitem{HalliwellHawking}
{
J. J. Halliwell and S. W. Hawking,
 ''Origin of structure in the universe'',
 {\it Phys. Rev. D} {\bf 31}, 1777 (1985). 
}

\bibitem{LindeInfl}
{
 A. Linde,
 ''Inflation and quantum cosmology'',
 {\it Physica Scripta T} {\bf 36}, 30 (1991).
}

\bibitem{Page}
{
 D. N. Page, ''Hawking's wave function of the universe'',
 {\it in}: R. Penrose and C. J. Isham (eds.),
 {\it Quantum concepts in space and time}, 
 Clarendon Press (Oxford, 1986), p. 274. 
}

\bibitem{Kiefer1}
{
 C. Kiefer,
 ''Wave packets in minisuperspace'',
 {\it Phys. Rev. D} {\bf 38}, 1761 (1988). 
}

\bibitem{Halliwell3}
{
 J. J. Halliwell,
 ''Introductory lectures on quantum cosmology'',
 {\it in}: S. Coleman {\it et. al.} (eds.),
 {\it Quantum cosmology and baby universes},
 World Scientific (Singapore, 1991), p. 159.
}

\bibitem{HawkingPage}
{
 S. W. Hawking and D. N. Page,
 ''Operator ordering and the flatness of the universe'',
 {\it Nucl. Phys. B} {\bf 264}, 185 (1986).
}

\bibitem{GradshteynRyzhik}
{
 I. S. Gradshteyn and I. M. Ryzhik, {\it Table of Integrals, Series
 and Products}, Academic Press, 1980, p. 1067. 
}

\bibitem{Bialynickietal}
{
 I. Bialynicki-Birula, M. Cieplak and J. Kaminski,
 {\it Theory of Quanta}, Oxford University Press, 1992, p. 286. 
}

\bibitem{FE8}
{
 F. Embacher,
 ''On the interpretation of quantum cosmology'',
 {\it preprint} UWThPh-1996-22, also gr-qc/9605019. 
}

\bibitem{Kiefersemi}
{
 C. Kiefer,
 ''The semiclassical approximation to quantum gravity'',
 {\it in}: J. Ehlers, H. Friedrich (eds.),
 {\it Canonical Gravity: From Classical to Quantum},
 Springer (Berlin, 1994), p. 170. 
}

\bibitem{Hawking1}
{
 S. W. Hawking,
 ''The path-integral approach to quantum gravity'',
 {\it in}: S. W. Hawking and W. Israel (eds.),
 {\it General relativity: An Einstein Centenary Survey},
 Cambridge University Press (Cambridge, 1979), p. 746. 
}

\bibitem{HartleHawking}
{
 J. B. Hartle and S. W. Hawking,
 ''Wave function of the universe'',
 {\it Phys. Rev. D} {\bf 28}, 2960 (1983).
}

\bibitem{Vilenkininter}
{
 A. Vilenkin,
 ''Interpretation of the wave function of the universe'',
 {\it Phys. Rev. D} {\bf 39}, 1116 (1989). 
}

\bibitem{GrishchukRozhansky}
{
 L. P. Grishchuk and L. V. Rozhansky,
 ''On the beginning and the end of classical evolution
 in quantum cosmology'',
 {\it Phys. Lett B} {\bf 208}, 369 (1988);
 ''Does the Hartle-Hawking wavefunction predict the
 universe we live in?'',
 {\it Phys. Lett B} {\bf 234}, 9 (1990). 
}

\bibitem{Lukas}
{
 A. Lukas,
 ''The no-Boundary wave function and the duration of the
 inflationary period'',
 {\it Phys. Lett B} {\bf 347}, 13 (1995). 
}

\bibitem{Zeh} 
{
 H. D. Zeh,
 ''Time in quantum gravity'',
 {\it Phys. Lett. A} {\bf 126}, 311 (1988).  
}

\bibitem{Lindeout}
{
 A. D. Linde,
 ''Quantum creation of the inflationary universe'',
 {\it Lett. Nouvo Cimento} {\bf 39}, 401 (1984). 
}

\bibitem{Vilenkintunnell}
{
 ''Boundary conditions in quantum cosmology'',
 {\it Phys. Rev. D} {\bf 33}, 3560 (1986). 
}

\bibitem{HosoyaMorikawa}
{
 A. Hosoya and M. Morikawa,
 ''Quantum field theory of the universe'',
 {\it Phys. Rev. D} {\bf 39}, 1123 (1989). 
}

\bibitem{McGuigan}
{
 M. McGuigan,
 ''Universe creation from the third-quantized vacuum'',
 {\it Phys. Rev. D} {\bf 39}, 2229 (1989). 
}

\end{thebibliography}
\end{document}